# Early theories on fluid resistance and translation of Euler's "*Dilucidationes de resistentia fluidorum*"

Sylvio R. Bistafa
sylvio.bistafa@gmail.com

## Abstract

In 1763, Euler published *Dilucidationes de resistentia fluidorum* (Explanations on the resistance of fluids), a memoir that challenges the fluid resistance theories proposed by Isaac Newton and d'Alembert. Euler's work explores the resistance experienced by solid bodies moving through fluids, critiquing both Newton's "common rule" and d'Alembert's paradox, which predicted zero resistance for non-viscous fluids. Euler's treatise is divided into two parts: the first focuses on the mathematical modeling of fluid flow patterns, while the second addresses the calculation of fluid resistance on surfaces. Despite significant advancements, Euler's work remains constrained by the limitations of non-viscous fluid assumptions, ultimately grappling with the same paradoxes he sought to overcome. This paper reviews the key contributions and limitations of *Dilucidationes*, emphasizing the ongoing relevance of Euler's insights in the context of classical fluid dynamics. Additionally, it highlights the translation capabilities of AI-powered tools, specifically ChatGPT, in translating complex mathematical texts, marking a noticeable improvement in handling notation-heavy content.

## 1. Introduction

In this memoir of 1763, *Dilucidationes de resistentia fluidorum* (Explanations on the resistance of fluids), Euler develops a new theory for the calculation of the resistance encountered by solid bodies moving through fluids. It begins by questioning the method proposed by Isaac Newton to calculate fluid resistance laid down in the *Principia* (1687, 1713) and referred to by Euler throughout his treatise as the "common rule" in which the resistance *"… is considered to follow a composite ratio, derived from the square of the velocity with which the fluid strikes and the square of the sine of the angle formed between the direction of the impact and the surface struck."* Nonetheless, Euler recognizes that it deserves to be recommended both for its ease of use in calculations and, most of all, because it is generally found not to deviate significantly from the truth.

Another motivation is that Euler was also critical of d'Alembert's work on the subject, by not been able to adequately explain the question of the resistance experienced by bodies moving in water and how it could be determined. Here he was referring to the so-called d'Alembert's paradox, which predicts zero resistance on a body moving with constant velocity relative to the fluid, in direct contradiction to the observation of substantial resistance on bodies moving relative to fluids. d'Alembert, working on a 1749 Prize Problem of the Berlin Academy on flow resistance, concluded: *"It seems to me that the theory, developed in all possible rigor, gives, at least in several cases, a strictly vanishing resistance, a singular paradox which I leave to future Geometers to elucidate"*.

Nonetheless, d'Alembert results were correct since he had modelled the non-viscous perfect fluid. Almost a century later, a first step towards solving the paradox was made by Saint-Venant, who instead had modelled the viscous real fluid, and in 1847 states: *"But one finds another result if, instead of an ideal fluid – object of the calculations of the geometers of the last century – one uses a real fluid, composed of a finite number of molecules and exerting in its state of motion unequal pressure forces or forces having components tangential to the surface elements through which they act; components to which we refer as the friction of the fluid, a name which has been given to them since Descartes and Newton until Venturi."*

It is not possible to visualize the flow of a non-viscous fluid because such a fluid does not exist. However, it is possible to compute the configuration of streamlines in the non-viscous flow over a flat plate, for example. Such a streamline configuration is shown in Fig. 1. It can be observed that only two streamlines actually touch the surface, and these streamlines are called stagnation streamlines. They divide the portion of the fluid that flows over the top of the plate from the portion that flows below it. We know that in regions where the streamlines are close to each other, the flow velocity increases and the pressure drops (Bernoulli's effect). Conversely, in regions where the streamlines move farther apart, the velocity decreases and the pressure increases. If we rotate this diagram by 180º, the streamline configuration remains unchanged (for any inclination of the plate relative to the incident flow). Since the pressure distribution is the same on both the upper and lower surfaces, the resultant pressure forces on the plate will be zero—there will be no resistance or lift. Without viscosity, the resultant resistance and lift will be zero for bodies of any shape, according to d'Alembert's paradox.

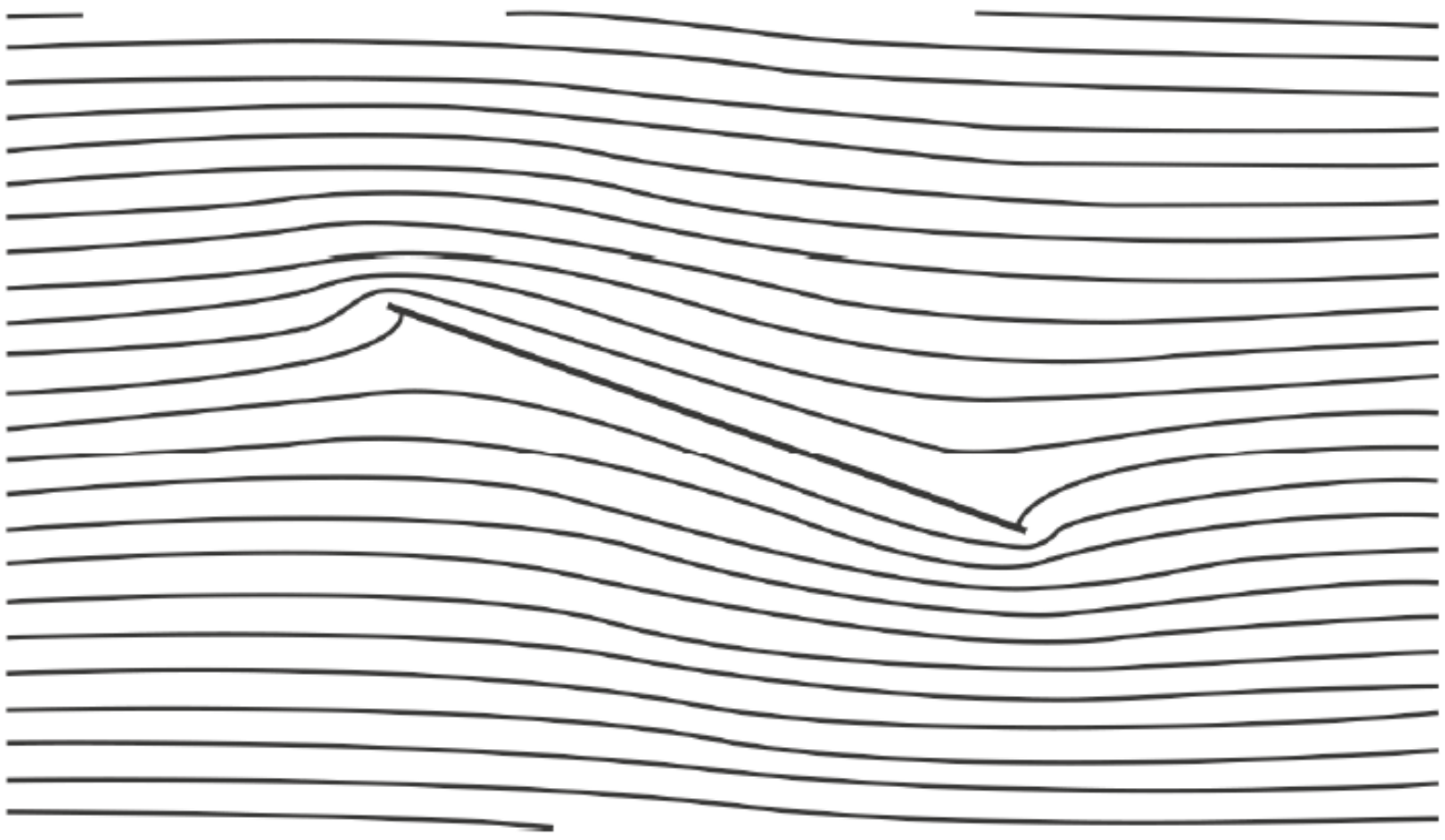

Figure 1: Non-viscous flow around a flat plate. There is no resistance or lift.

To help us understand the different behavior of the flow around a body according to the fluid characteristics, Fig. 2 shows the configuration of the streamlines of the flow around a circular cylinder of a perfect fluid (non-viscous) and a real fluid (viscous). In the flow of the perfect fluid (Fig. 2a) around the circular cylinder, symmetry in the configuration of the streamlines is observed, both with respect to the x-axis and the y-axis. The symmetry of the streamlines also implies symmetry in the pressure distribution with respect to both axes. In the case of the real fluid flow around the circular cylinder (Fig. 2b), the pressure distribution on the front half of the cylinder is different from that on the rear half, as the symmetry of the streamline configuration is no longer preserved with respect to the y-axis. However, since the streamline configuration remains symmetrical with respect to the x-axis, the pressure distribution on the upper half is the same as on the lower half of the cylinder. The asymmetry in the pressure distribution with respect to the y-axis gives rise to resistance on the cylinder.

Naturally, this understanding was not available at the time. Theories developed without considering the effects of viscosity, ultimately led to the formulation of d'Alembert's paradox. As we shall see, this was also the case of Euler's attempt in *Delucidationes*.

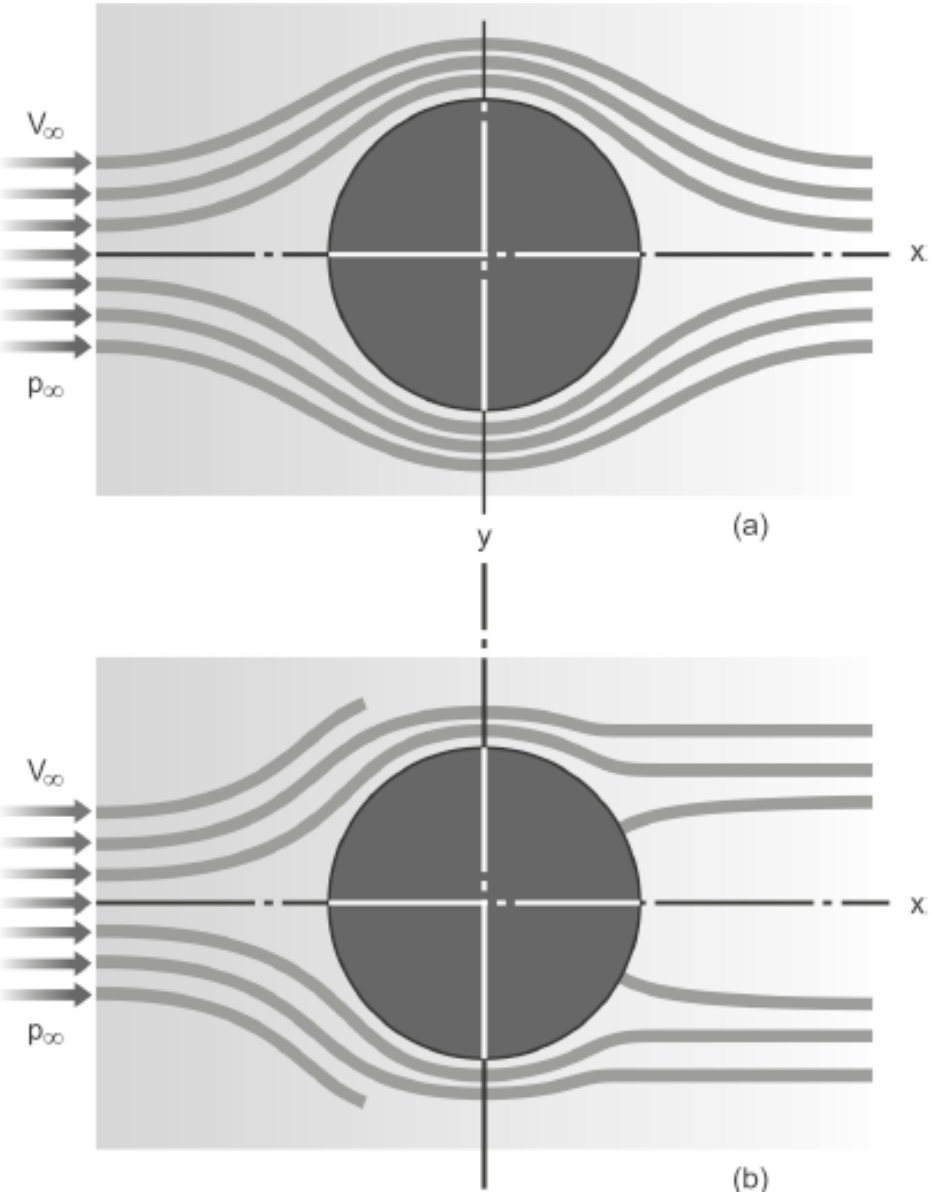

Figure 2: Configuration of the streamlines for the flow around the circular cylinder: perfect fluid (a); real fluid (b).

## 2. General Overview of *Delucidationes*

*Delucidationes* is a comprehensive treatise consisting of 48 paragraphs, which can be divided into two parts. In the first part, spanning approximately 24 paragraphs, Euler mathematically describes the flow patterns around bodies using streamlines, with a particular focus on flows around ships. The final 24 paragraphs of *Delucidationes* are dedicated to the crux of the subject: a method for calculating the pressures around the surface of a body, from which resistance can be determined.

We now present a general overview of Euler's *Delucidationes*—a significant work that delves into the complex phenomena of fluid resistance. This treatise not only builds upon earlier theories but also attempts to reconcile the mathematical intricacies with physical observations. Through an examination of key points put forth in the *Delucidationes*, we aim to provide a clearer understanding of its contributions and limitations to the topic.

In §IV, Euler criticizes the common rule for calculating the force exerted by the fluid on the body, arguing that it inaccurately assumes a direct impact of the fluid on the body. Instead, the author explains that the fluid currents change their course as they approach the body, flowing along its sides rather than colliding directly with it (Fig. 3). He emphasizes that the force exerted by water, referred to as resistance, is primarily due to the pressure of the water flowing past the body.

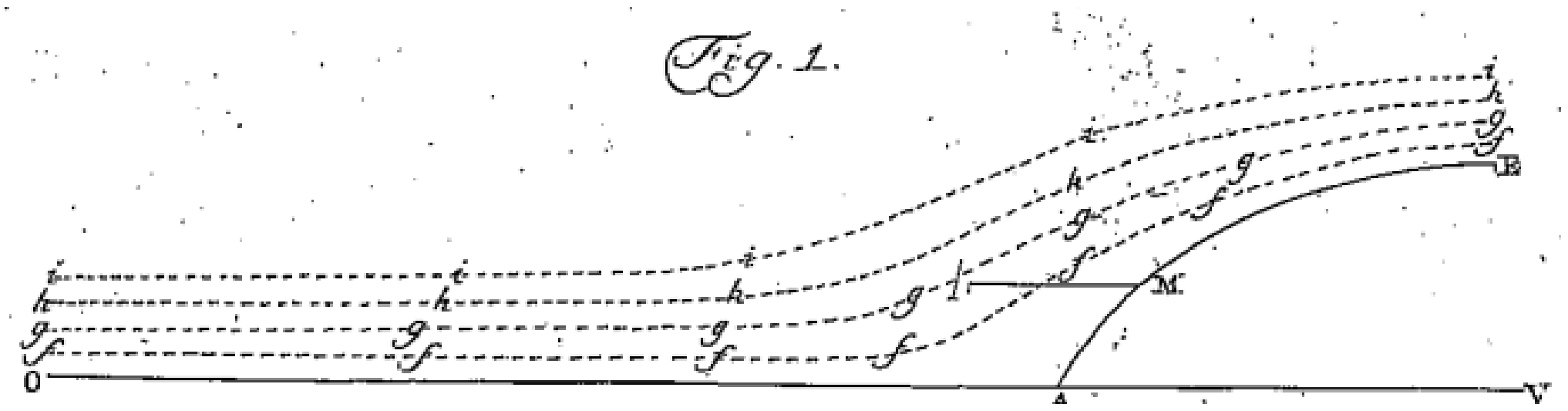

Figure 3: Euler's sketch illustrating the configuration of streamlines around the body AMB. This is likely one of the earliest clear depiction of streamlines around a body in the history of fluid dynamics literature.

In §V, Euler argues that to accurately determine resistance, one must investigate this pressure, which involves a complex analysis of the fluid's motion around the body.

In §VI, Euler reflects on the challenges faced by mathematicians and scientists, including the renowned d'Alembert, in accurately determining the resistance that fluids exert on moving bodies.

In §VIII, Euler explains how to calculate the pressure exerted by water on a body at a specific point M. The velocity of the water at M is assumed to be related to a certain height $v$, and the pressure at M is given by the difference $k - v$, where $k$ is a constant for the entire stream. This pressure is understood as the force exerted by a column of water with height $k - v$ on the body's surface at M.

In §IX, Euler discusses the relationship between fluid velocity and resistance. The author points out that, contrary to what might be expected, increased velocity in a stream results in decreased pressure, and thus decreased resistance, while decreased velocity results in increased pressure and resistance.

In §XI, Euler discusses how the common rule, despite its imperfect foundation, can be used to approximate the velocity of water as it flows past a body. By recognizing that the amplitude of the stream is inversely proportional to the velocity, the amplitude of the water touching the body can be determined at each point.

In §XII, Euler develops an expression for the velocity at a given point on the surface of a body. For the flow around the body AMI, the resistance according to the common rule is given by $c\, sin^2\, \phi$, where $\phi$ is the angle of incidence AMI, formed by drawing a straight line MI parallel to the direction AO (see Fig. 3), and $c$ is the dynamic pressure (velocity squared) of the uniform stream far away from the body. This allows for the calculation of the actual velocity of the water at each point on the body's surface.

In the next paragraphs, Euler diverges to such questions as: how to estimate the speed of a ship by observing the velocity of the water flowing past it (§XIV), design considerations for the stern of a ship (§ XVI), where all efforts should focus on avoiding the other issue, where resistance significantly increases due to the near-vacuum left behind the ship. Here Euler warns on the possibility of cavitation, arising from negative pressures, which is a pioneering and notable remark.

In §XVII, Euler explores how the shape of the streams (or rivulets) near a ship's body can be derived from the velocity of the water as it flows past. By understanding that the stream width is inversely proportional to the water's velocity at each point (law of continuity), Euler provides a method to determine the curvature of the streams. The mathematical relationships between the velocity, angles, and distances help construct the curve that represents the flow pattern around the ship's body.

In §XVIII – §XXIII, Euler provides a detailed geometric analysis of how the width and characteristics of the fluid stream change as it flows around a body, with a focus on the behavior of water near the surface and perpendicular to the current. These passages collectively refine the understanding of fluid flow around a body by analyzing the flow on simple cases: an inclined surface AE, and a horizontal obstacle BE, highlighting variations in resistance as the water moves around (Fig. 4).

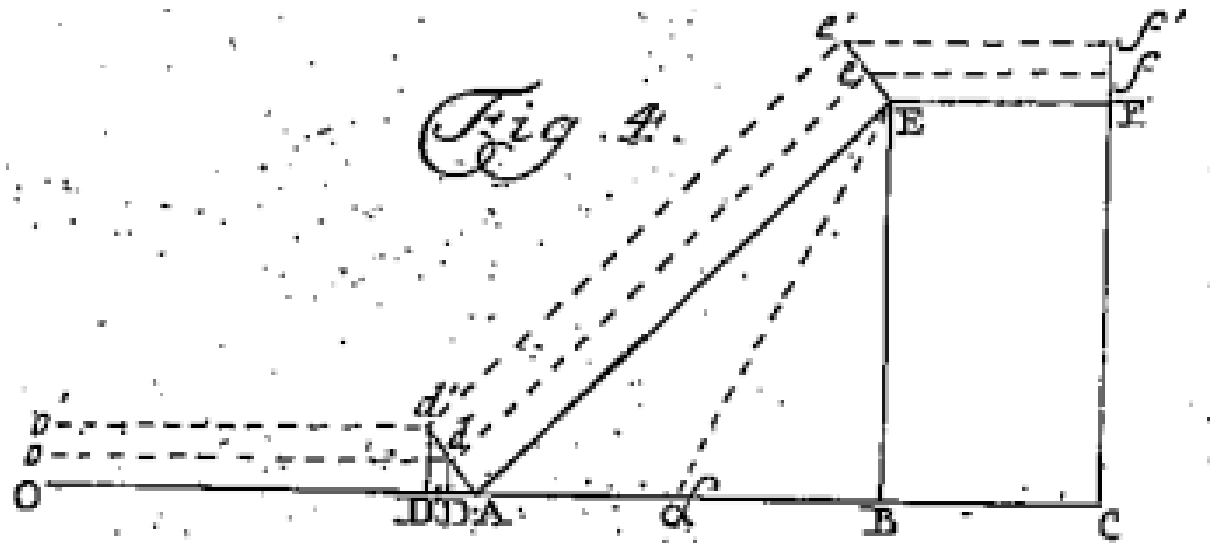

Figure 4: Simple cases used by Euler for the analysis of the flow on an inclined surface AE and on a horizontal obstacle BE.

In §XXIV, Euler explores the geometry of water flow around a body shaped like a quarter-circle (Fig. 5). By setting up equations that describe the curves of the water streams, the author illustrates how the shape of the body influences the water’s trajectory and the resulting flow patterns. The passage also hints at the complexity of these curves, which can vary significantly based on their distance from the body.

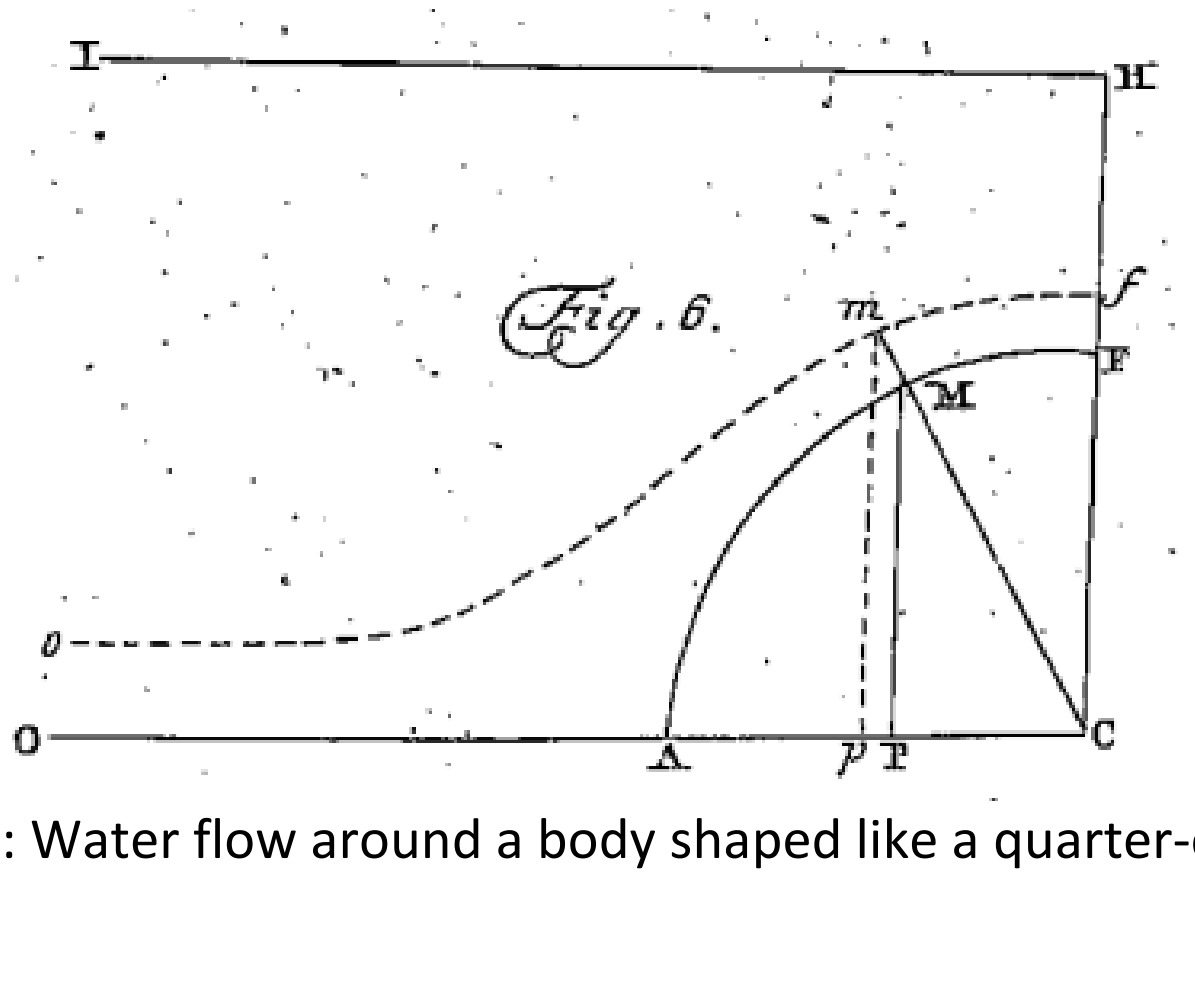

Figure 5: Water flow around a body shaped like a quarter-circle.

In §XXV, Euler generalizes the previous findings to accommodate any curve, not just a specific geometric shape like a quarter-circle. The solution is expressed in terms of infinite series, allowing for the computation of the stream paths for more complex body shapes. The complexity of the method led Euler to a dead end.

In §XXVII, Euler introduces a more complex scenario where the body is placed in a confined fluid channel, and the fluid must flow around the body within the given space. Since in this case the standard rule does not apply, the velocity is determined based on the body's geometry and the channel's dimensions, which is given by:

$$v = \frac{cbbdx^2}{(b-y)^2(dx^2+dy^2)},$$

where $c$ is the body velocity and $b$ is the channel width.

In §XXVIII, Euler presents a mathematical formulation to calculate the resistance experienced by a body moving through a fluid within a confined channel. The approach relies on the "true theory of resistance", which ties the pressure at a point on the body's surface to the velocity of the fluid at that point. The pressure distribution is used to derive an expression for the resistance, considering both finite and infinite fluid scenarios. Key points of the analysis include:

1. **Pressure Calculation**: The pressure at a point M on the surface of the body is determined from the velocity at that point, and it is given by:

   $$-\frac{cbbdy}{(b-a)^2} + \frac{cbbdx^2dy}{(b-y)^2(dx^2+dy^2)}.$$

2. **Resistance Formula**: The pressure formula is then used to derive the resistance on an infinitesimal element of the body's surface. By integrating this expression, the total resistance can be calculated. The results show that the resistance depends on both the finite and infinite nature of the fluid. Integrating this expression will yield the total resistance. If the fluid's amplitude $b$ were infinite, the resistance would be:

   $$= -cy + c\int \frac{dx^2dy}{dx^2+dy^2}.$$

   Now, if AMF is a straight line AF, and $CA = b$, with $CF = a$, then the resistance $R$ would be:

   $$R = \frac{a^3c}{aa+bb}.$$

3. **Infinite vs. Finite Fluid**: The resistance in an infinite fluid is simpler and serves as a reference value $R$. However, in a finite fluid scenario (such as within a channel), the resistance is shown to be greater, reflecting the constraints imposed by the channel, and it would be given by:

   $$\frac{b(ab+bb)}{a(b-a)^2}R.$$

In §XXXI, Euler applies the principles of fluid resistance to the design of ships, and in §XXXII, Euler addresses the importance of stern design and its impact on both rudder effectiveness and overall ship resistance.

In §XXXIV, Euler introduces the concept of orthogonal trajectories (Fig. 6), which intersect the streamlines at right angles, to better understand the fluid's behavior around the body. The streamlines and the orthogonal coordinates form a mathematical framework to describe the position of points along streamlines in a fluid flow, which are known as *intrinsic coordinates*. Euler pioneered the use of these coordinates, which represents a revolutionary approach to the analysis of fluid flow.

In §XXXV, Euler introduces coordinates $s$ and $b$, where $s$ represents the abscissa of a point on the curve AME and $b$ is a parameter that distinguishes different streamlines. These are the intrinsic coordinates on the body's surface. The differential equations describe how the orthogonal coordinates $x$ and $y$ change with respect to the intrinsic coordinates $s$ and $b$.

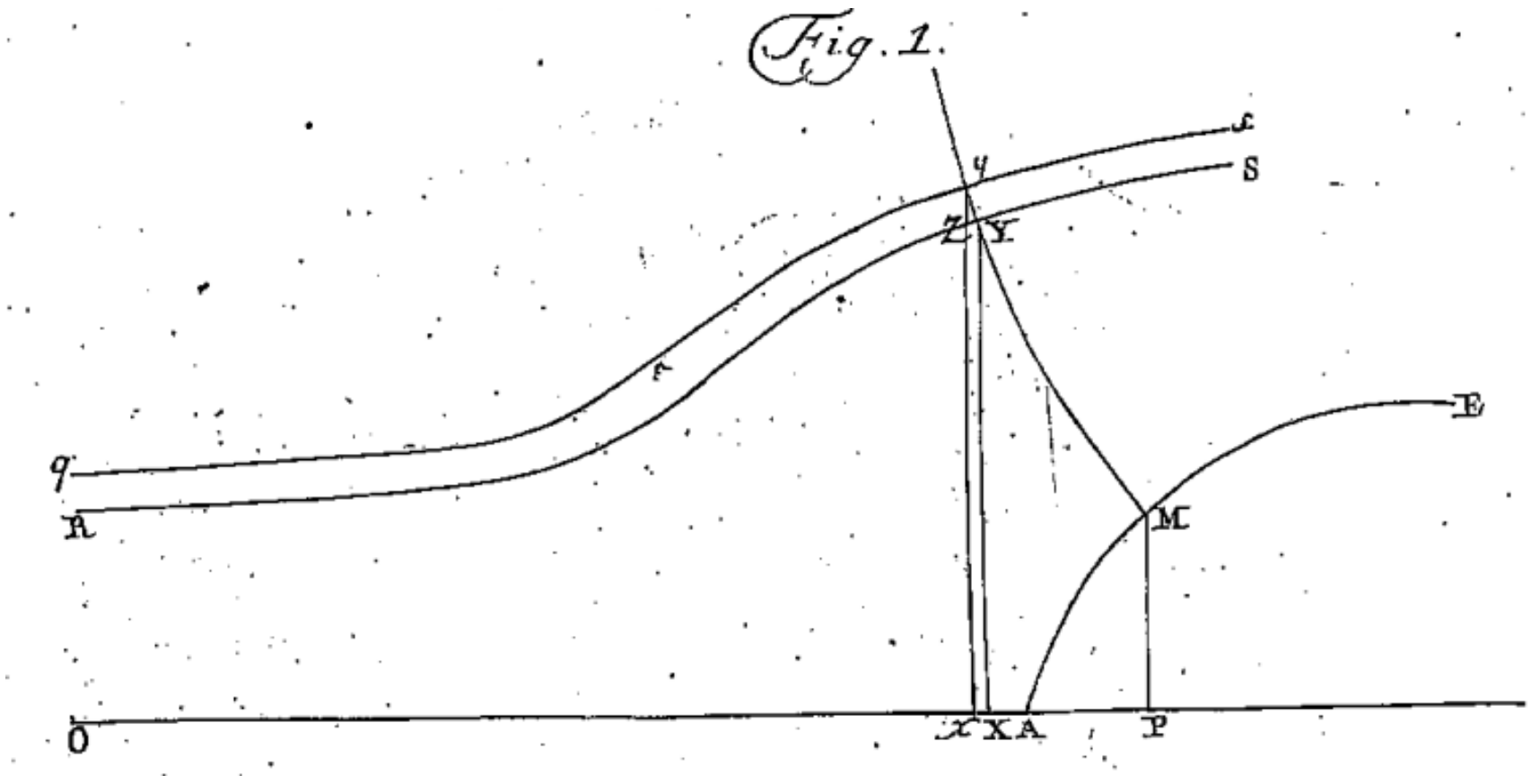

Figure 6: Intrinsic coordinates: streamline RS, orthogonal trajectory MYy (not a fluid trajectory).

In §XXXVII, Euler continues developing the mathematical framework for fluid dynamics by expressing velocity components $u$ and $v$ in terms of other variables and ensuring that the resulting formulas are integrable. He introduces specific conditions to satisfy the physical and mathematical consistency of the theory, refining the equations governing fluid motion.

In §XXXVIII, Euler gives an equation for the calculation of the pressure $p$ as:

$$p = V - \int udx\frac{du}{dx} + vdx\frac{du}{dy} + udy\frac{dv}{dx} + vdy\frac{du}{dy},$$

which for a single stream reduces to:

$$p = V - \frac{1}{2}88,$$

where 8 is the local velocity which is related to its components $u$ and $v$ by $uu + vv = 88$.

The expression given for $p$ can be seen as a result of integrating the momentum equation in two dimensions. The terms in the integral represent the contribution of the velocity gradients to the pressure field in the fluid. The term $V$ is the total potential or energy in the fluid, while the integral accounts for the kinetic energy contributions from the velocity components $u$ and $v$. The detailed steps for the derivation would involve setting up the Euler equations, integrating them with respect to the spatial variables $x$ and $y$, and then expressing the pressure $p$ as a function of the velocity components and their derivatives. Euler likely used the method of integration by parts or similar techniques to arrive at this final expression.

Because of the complexity of the mathematical developments alluded to earlier, we now move directly to §XLIII - §XLVII, where Euler presents his sole application. In these paragraphs, Euler addresses an inverse problem—identifying a body shape for which the "common rule" holds true, without acknowledging that Newton's resistance formula actually pertains to total resistance rather than local resistance. The main results of this section are the equations governing the streamlines and pressure, as given below:

$$x^2 = \frac{4}{C^2}\left(\frac{1}{C^2} - y\right);\; p = V - \frac{\frac{2c^2}{C^2}}{x^2 - \frac{2c^2}{C^2}},$$

where $c$ is another constant.

In §XLVIII, Euler concludes that the common rule of fluid resistance is only accurate under very specific conditions—when the body and the fluid streams follow a parabolic shape. “… Consequently, the outer edge of the vessel αεβ must also be shaped like a similar parabola. In all other cases, the rule does not apply, and resistance behaves differently. Since the parabolic body AEB [Fig. 7] is not terminated on both sides, this scenario cannot be applied to the theory of resistance. Consequently, when some were misled by this rule and believed that a body could move through a fluid without experiencing resistance [based on the idea that the fluid's action on the rear cancels out the force on the front] cannot hold in the real world, because this cancellation is hindered by viscosity.”

A parabolic shape that adheres to the common rule of resistance refers to a specific shape of a body, typically a parabolic curve, that follows the "common rule” regarding how resistance is calculated. When a body has a parabolic shape, this geometry might align with this rule, leading to predictable and often minimal resistance under ideal conditions (such as in a non-viscous fluid). This adherence to the rule implies that the parabolic shape is optimal or ideal in the context of the simplified assumptions of the model.

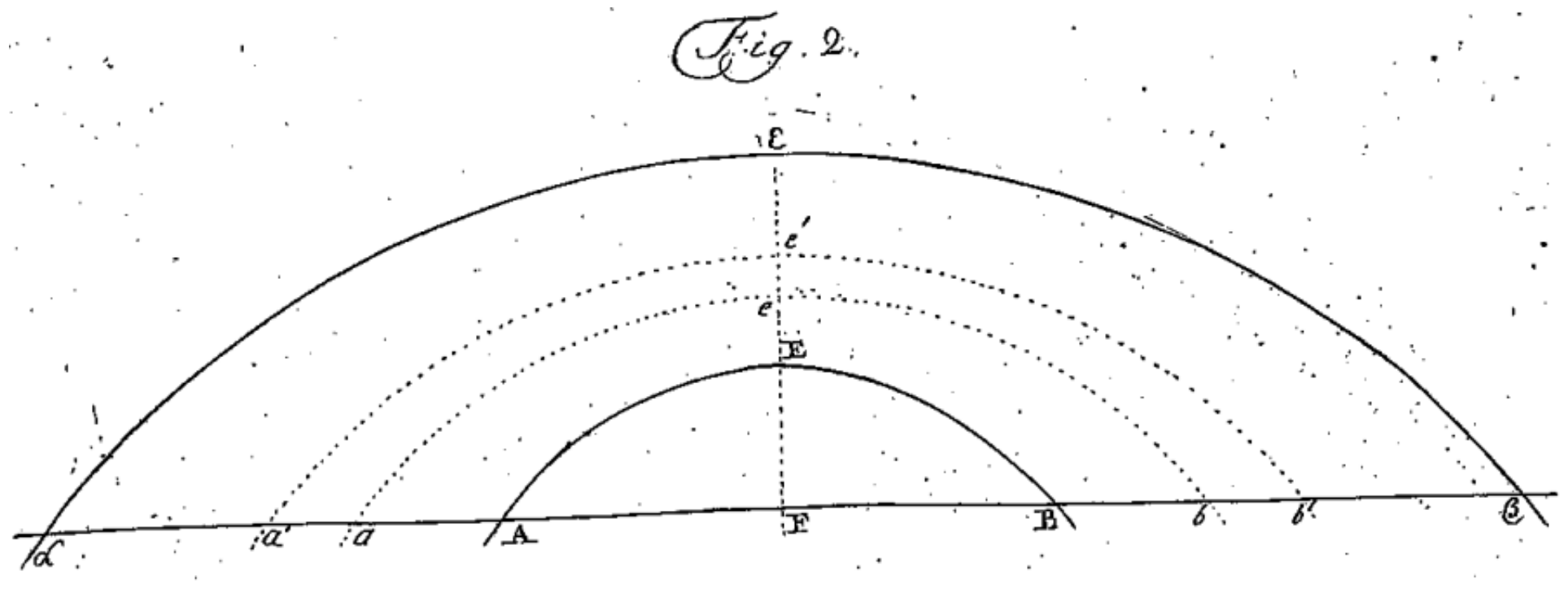

Figure 7: Parabolic shape that adheres to the common rule of resistance.

Euler's recognition at the end that the common rule of resistance is only valid under very specific conditions, and his acknowledgment that the general case deviates from this rule, implies a recognition of the limitations inherent in ignoring viscosity. The d'Alembert paradox, which states that a non-viscous fluid cannot exert resistance on a body moving through it, essentially highlights the necessity of considering viscosity to explain real-world fluid resistance.

By acknowledging that the rule of resistance fails when viscosity is ignored, Euler indirectly concedes that the paradox holds in the idealized case of non-viscous flow. This indicates that without considering the effects of viscosity, the theoretical predictions of resistance in a fluid will not match the observed reality, as the paradox suggests.

## 3. The translation

In a previous paper[1], the author observed that the AI-powered chatbot, ChatGPT, exhibited a strong proficiency in translating Latin. However, the author also noted the importance of caution, particularly when translating mathematical content that is dense with notation.

In this current study, the author extended the testing of ChatGPT by translating a 32-page scientific text by Euler. This text develops a mathematical model for a classical fluid dynamics problem, specifically the calculation of resistance encountered by solid bodies moving through fluids.

During the translation, the author noticed a marked improvement in ChatGPT's ability to handle complex mathematical texts compared to the earlier exercise. Notably, no specific recommendations or corrections were required for the initial translations provided by ChatGPT, which was particularly impressive. Additionally, without prompting, ChatGPT offered insightful comments after each of the 48 paragraphs inputted from the original publication.

Intrigued by this enhanced capability in translating notation-heavy mathematical Latin texts, the author posed the following question to ChatGPT: "I have noticed a significant improvement of ChatGPT in translating a notation-heavy mathematical Latin text compared to a previous experience with this chatbot on similar texts. Is it possible that the free version of ChatGPT available online has automatically improved its translating capabilities in a short span of time?" The response provided by ChatGPT is presented next.

*"Firstly, you are interacting with GPT-4, which has a knowledge cutoff in September 2023:*

1. ***Model Updates**: The capabilities of ChatGPT, like any AI, can be improved over time with new updates. If you've noticed an improvement in a short span, it's possible that the version you're interacting with has been updated. OpenAI periodically releases new versions of its models, which can include enhancements in language understanding, including translating complex texts with specialized notations.*

[1] Translating scientific Latin texts with artificial intelligence: the works of Euler and Contemporaries. To appear in *Euleriana*.

2. ***Enhanced Contextual Understanding**: The model might have become better at handling complex texts by improving its ability to maintain and understand context. As more users interact with the system and more diverse data is incorporated, the AI's ability to process challenging texts, such as mathematical Latin, can improve.*
3. ***Training and Fine-Tuning**: AI models like ChatGPT are often fine-tuned on specific datasets to improve their performance on tasks like translation. The improvements you're seeing could result from such targeted fine-tuning, which enhances the model's ability to deal with particular challenges like notation-heavy texts.*
4. ***Dynamic Learning**: Even in a free version, there might be mechanisms that allow the model to adapt better to certain types of queries based on feedback and interaction patterns over time.*

*While these improvements may not happen in real time, they can be periodically integrated into the versions of the model that are made publicly available, leading to noticeable changes in performance even in short spans of time. So, while it's difficult to pinpoint exactly what led to the improvement you're observing without knowing the exact changes made, it's quite possible that updates to the model have enhanced its translating capabilities."*

## 4. Conclusions

Euler's *Dilucidationes de resistentia fluidorum* represents a significant yet complex contribution to fluid dynamics, challenging the prevailing theories of his time. While Euler advanced the understanding of fluid resistance and flow patterns, his work was ultimately constrained by the same paradoxes that hindered his contemporaries, particularly the limitations of modeling non-viscous fluids. Euler's attempt to reconcile mathematical theory with physical observation, though intricate and groundbreaking, could not fully resolve the issue of resistance in real-world scenarios where viscosity plays a critical role.

The translation of Euler's treatise using ChatGPT demonstrated remarkable proficiency in handling notation-heavy mathematical Latin texts. The AI tool showed substantial improvement over previous versions, requiring no manual corrections and even offering insightful comments on the text. This suggests that advancements in AI translation capabilities can significantly enhance the accessibility of historical scientific works, facilitating deeper engagement with foundational texts in fields like fluid dynamics.

Overall, Euler's *Dilucidationes* remains a foundational yet challenging text that continues to inspire inquiry into the complexities of fluid resistance, while the use of AI tools like ChatGPT opens new avenues for exploring and understanding such historical contributions.

## Notes

The translation presented here and some texts provided have been generated with the assistance of AI technology.

OpenAI, the organization behind ChatGPT, recommends providing appropriate context and verifying the information from reliable sources whenever possible, as AI-generated responses should be considered as a tool to assist and augment human knowledge rather than a definitive source of information.

The present interaction with GPT-4, has a knowledge cutoff in September 2023.

## Declarations

No funding was received for conducting this study.

The author has no competing interests to declare that are relevant to the content of this article.

Explanations on the resistance of fluids
E276 -- *Dilucidationes de resistentia fluidorum*
*Auctore*
*L. Eulero*

Translated and Commented
by
ChatGPT
September 2024

*I.*

*Duplici modo quaestio de resistentia, quam corpora solida in fluidis mota patiuntur, tractari solet, altero negotium tantum vero proxime plerumque conficitur, dum quantitas resistentiae per regulam satis concinnam ad calculum revocatur; altero vero resistentiae doctrinam ex ipsa fluidorum natura et pressione, quam in corpora exerunt, per profundíssimas Hydrodynamicae. inuestigationes constituere Geometrae sunt conati. Quo posteriori modo si negotium ad finem perducer liceret, omnia, quae ad mensuram resistentiae pertinent, inde accuratissime definiri possent, neque amplius coacti essemus, ad modum priorem confugere, quo prope tantum vera resistentiae magnitudo exhibetur. Verum etiam nunc tam longe ab ista perfecta resistentiae cognitione abesse videmur, ut priori modo, etiamsi eius defectum probe norimus, minime carere queamus, sed eo potius, quoties resistentia indaganda occurrit, uti debeamus.*

I.

"A question concerning the resistance that solid bodies experience when moving in fluids is usually addressed in two ways. In the first method, the task is generally completed more or less accurately, as the amount of resistance is calculated using a fairly convenient rule. In the second method, however, mathematicians have attempted to establish the theory of resistance based on the very nature of fluids and the pressure they exert on bodies, through the most profound investigations of Hydrodynamics. If it were possible to carry this task to its conclusion in this latter manner, all aspects related to the measurement of resistance could be defined with the utmost accuracy, and we would no longer be compelled to resort to the former method, which only provides an approximation of the true magnitude of resistance. However, even now, we seem to be so far from this perfect understanding of resistance that we cannot do without the first method, even though we are well aware of its shortcomings. On the contrary, we must use it whenever the investigation of resistance arises."

This passage reflects a discussion on the methods used in understanding and calculating the resistance encountered by solid bodies moving through fluids, highlighting the limitations of current knowledge and the challenges in achieving a perfect understanding of this phenomenon.

*II. Prior autem modus, quo Newtonus plurinum est usus, etiamsi eius aberrationem a veritate non ignorasse videatur, hac regula ad calculum inprimis accommodata continetur, ut resistentia rationem compositam sequi censeatur, ex ratione duplicata celeritatis, qua fluidum impingit, et ratione pariter dueplicata sinus anguli, quem directio impulsionis cum superficie percussa constituit. Hinc ergo si pro allisione fluidi perpendiculari, ubi angulus ille sit rectus, resistentiae quantitatem noverimus, facile erit, eam pro quauis allisione obliqua assignare. At vero si fluidum perpendiculariter superficiem quampiam planam feriat, resistentia aequalis aestimatur ponderi columnae eiusdem fluidi, cuius basis sit ipsa superficies percussa, altitudo vero congruat cum ea, ex qua grave cadendo ipsam fluidi celeritatems esset impetraturum.*

"II. The first method, which Newton used extensively, even though he seems not to have been unaware of its deviation from the truth, is particularly suited to calculation by this rule: that resistance is considered to follow a composite ratio, derived from the square of the velocity with which the fluid strikes and the square of the sine of the angle formed between the direction of the impact and the surface struck. Therefore, if we know the amount of resistance for a perpendicular impact, where that angle is right, it will be easy to assign it for any oblique impact. But if a fluid strikes a certain flat surface perpendicularly, the resistance is estimated to be equal to the weight of a column of the same fluid, whose base is the struck surface itself, and whose height corresponds to that from which a heavy body, falling, would acquire the same velocity as the fluid."

This passage discusses the method used by Isaac Newton to calculate fluid resistance, noting that while Newton was aware of the method's limitations, it is nonetheless useful for practical calculations. The passage explains how resistance can be calculated based on the velocity of the fluid and the angle of impact, and it provides a specific rule for estimating resistance when a fluid strikes a flat surface perpendicularly.

*III. Haec regula cum ob facilem vium in calculo, tum vero ideo potissimum commendari meretur, quod a veritate plerumque haud notabiliter abludere deprehendatur. Nam quod ad principia attinet, quibus innititur, nullum plane est dubium, quin ea nimis sint vaga, atque a vero statu, ad quem accommodantur, remota, quam ut conclusio inde deducta pro certa admitti queat. Maximam enim partem haec regula est petita ex collisione corporum, dum fluidum continuo in corpus data celeritate et secundum directionem motus sui impingere, conflictumque exerere conciptur, At vero certum est, fluidum neutiquam in corpus hoc modo impingere , sed antequam ad corpus perveniat, tam suam directionem, quam celeritatem, ita inflectere, ut cum ad corpus pervenerit, secundum ipfam eius superficiem praeterlabatur, nullamque aliam vim in corpus exerat, praeter pressionem, quae ipsi in singulis contactus punctis convenit. Quam ob rem counclusio, quae ex ratiocinio tam perverso deduci solet, minime pro vera haberi potest.*

"III. This rule, although it deserves to be recommended both for its ease of use in calculations and, most of all, because it is generally found not to deviate significantly from the truth, is based on principles that are too vague and distant from the actual conditions to which they are applied to be accepted as certain. For the most part, this rule is derived from the collision of solid bodies, assuming that a fluid continuously strikes a body with a given velocity and in the direction of its motion, thereby exerting an impact. But it is certain that a fluid does not strike a body in this manner; instead, before it reaches the body, it alters both its direction and velocity in such a way that, when it finally arrives at the body, it flows past it along the surface and exerts no other force on the body except the pressure appropriate to each point of contact. Therefore, a conclusion derived from such a flawed reasoning cannot be considered true."

This passage critiques the rule used to calculate fluid resistance, arguing that it is based on principles that are too imprecise and do not accurately reflect the true nature of fluid behavior. The author points out that the rule is derived from solid body collisions, which do not accurately model how fluids interact with surfaces. The fluid, rather than directly colliding with a body, adjusts its motion to flow along the surface, exerting only pressure. As such, conclusions drawn from this rule should not be regarded as true.

*IV. Quo hoc clarius perspiciamus, flumen concipamus, quod data celeritate fecundum directionem OV feratur; iam vero in hoc flumine corpus collocari AME, quod quantam vim a flumine sit sustentaturum, definiri oporteat. Atque per regulam vulgarem haec vis ita investigatur, quasi in singula corporis puncta M vena aquea IM fecundum directionem fluminis, eaque celeritate, qua flumen progredi assumimus, incurreret, ac per conflictum verum corpori vim inferret. Interim tamen si actionem fluminis, prouti revera se habet, perpendamus, mox percipiemus, tractus fev quasi rivulos fluminis, qui fupra corpus in notabili distantia celeritatem fuam cum: directione retinuerant, uti f, f, f, g, g, g etc. cum propius ad corpus accefierint; cursum suum inflectere, atque tandem iuxta corporis latera defluere, quae deflexio in figura exhibetur. Ex quo manifestum est, nusquam eiusmodi conflictum fieri, qualis in constitutione regulae vulgaris concipi folet.*

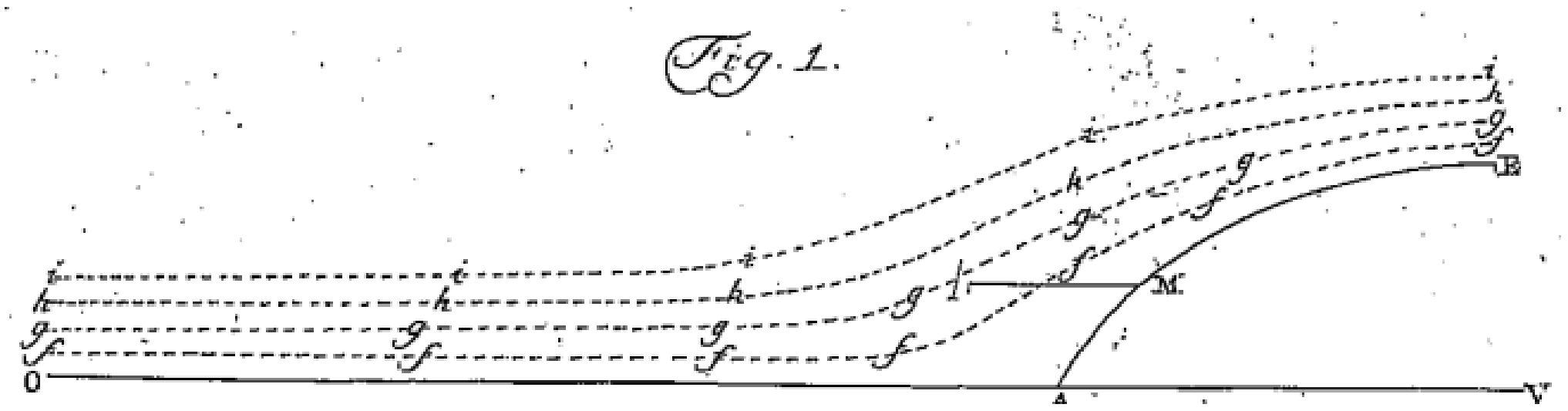

"IV. To make this clearer, let us imagine a stream that flows with a given velocity in the direction OV. Now, in this stream, a body AME is placed, and it is necessary to determine the force that the stream will exert on the body. According to the common rule, this force is investigated as if at each point M of the body, a stream of water IM, moving in the direction of the flow and with the assumed velocity of the stream, would strike it and exert a force through actual impact. However, if we consider the action of the stream as it truly occurs, we will soon perceive that the currents or small streams of the fluid, which retain their velocity and direction at a considerable distance from the body, such as f, f, f, g, g, g, etc., when they approach the body, change their course and eventually flow along the sides of the body, as depicted in the figure. From this, it is evident that no such impact occurs as is typically conceived in the establishment of the common rule."

This passage discusses the dynamics of a fluid stream interacting with a solid body placed within it. The author critiques the common rule for calculating the force exerted by the fluid on the body, arguing that it inaccurately assumes a direct impact of the fluid on the body. Instead, the author explains that the fluid currents change their course as they approach the body, flowing along its sides rather than colliding directly with it. This observation highlights a significant flaw in the traditional method of calculating fluid resistance.

*V. Quin potius hinc manifestum est, istam aquae vim, quae sub resistentiae nomine comprehenditur, a pressione aquae iuxta corpus praeterlabentis proficisci, quam idcirco pressionem investigari necesse est, si resistentiam accurate assignare velimus. Quare vera ratio resistentiam determinandi, qua alter modus supra memoratus continetur, huc redit, ut pressionem quam corpus in singulis punctis a fluido sustinet, definiamus: at vero haec quaestio altioris est indaginis, quam ut eius enodationem a profectibus, quos adhuc in hydrodynamicis fecimus, expectare queamus. Hic enim singuli rivuli, ex quibus fluvius constat, et quemadmodum cursum suum circa corpus inflectant, considerari, atque omnes illae lineae curvae ff, gg, bb etc. quafi sub communem aequationem redigi debent; unde deinceps aquae celeritas in singulis cuiusque rivuli punctis concludi queat. Hac autem demum celeritate cognita, ipsam pressionem, cui hoc negotium innititur, assignare licebit, a tam perfecta autem motus fluidorum cognitione adhuc longe absumus.*

"V. Rather, it is evident from this that the force of the water, which is understood under the name of resistance, arises from the pressure of the water flowing past the body. Therefore, it is necessary to investigate this pressure if we wish to accurately determine the resistance. Thus, the true method of determining resistance, as contained in the second method mentioned above, comes down to defining the pressure that the body sustains at each point from the fluid. However, this question requires a deeper investigation than what can be resolved by the advancements we have made so far in hydrodynamics. For here, each of the small streams that make up the river, and how they bend their course around the body, must be considered, and all those curved lines ff, gg, bb, etc., must be reduced, as it were, to a common equation, from which the velocity of the water at each point of each stream can subsequently be determined. Only after this velocity is known can the pressure, upon which the problem depends, be assigned. Yet, we are still far from such a perfect understanding of fluid motion."

This passage emphasizes that the force exerted by water, referred to as resistance, is primarily due to the pressure of the water flowing past the body. The author argues that to accurately determine resistance, one must investigate this pressure, which involves a complex analysis of the fluid's motion around the body. The passage highlights the challenges in hydrodynamics, noting that the current understanding of fluid motion is still insufficient to solve this problem perfectly.

*VI. Quae Celeb. Alembertus de resistentia fluidorum in peculiari Tractu est commentatus, hanc summam difficultatem, veram resistentiam investigandi, magis demonstrant, quam levant. Cum enim Vir acutissimus omni adhibita sagacitate hanc quaestionem adaequate explicare haud valuerit, ut inde resistentia, quam quaeuis corpora in aqua mota patiuntur, assignari possit: magno hoc nobis est argumento, quaestionem tantopere esse difficilem, ut vires humanas tantum non superare videatur. Quae ego etiam nuper in aliquot dissertationitus de motu fluidorum exposui, nullum subsidium huc afferunt. Etiamsi enim omnia, quae ad motum fluidorum pertinent, ad aequationes analyticas reduxi, tamen ipsa Analysis minime adhuc ita est exculta, ut illis aequationibus resolvendis sufficiat. Quae porro alii de hoc argumento sunt meditati, haud feliciori successu vires suas ingenii sunt experti.*

"VI. What the renowned d'Alembert has discussed concerning fluid resistance in a particular treatise demonstrates this great difficulty of accurately investigating true resistance more than it alleviates it. For even though this most acute man employed all his sagacity, he was not able to adequately explain this question so that the resistance experienced by bodies moving in water could be determined. This serves as a strong indication for us that the question is so difficult that it seems almost to exceed human capabilities. What I myself have recently set forth in several dissertations on fluid motion provides no assistance here. For although I have reduced everything related to fluid motion to analytical equations, yet Analysis itself is not yet sufficiently advanced to solve those equations. Furthermore, others who have meditated on this subject have not tested their intellectual powers with any greater success."

This passage reflects on the challenges faced by mathematicians and scientists, including the renowned d'Alembert, in accurately determining the resistance that fluids exert on moving bodies. Despite their best efforts and the use of sophisticated analytical methods, the problem remains so difficult that it seems almost beyond human capability to solve. The author also acknowledges that his own work in this area has not provided a solution, as the necessary analytical tools are not yet fully developed. Moreover, others who have attempted to address this issue have not been more successful.

*VII. Etsi autem determinatio pressionis in genere, hoc est in omnibus punctis fluidi, tam a tractu singuloram rivulorum, quam ab aquae celeritate pendet, tamen inveni, si quaestio ad unicum rivulum restingatur, tum pressionem in singulis eius locis per solam celeritatem definiri. Quare cum corpis AME ab unico rivulo f, f, f contingatur, omnisque resistentia ab eius pressionibus solis oriatur, si modo pressionem huius rivuli in singulis eius punctis cognosceremus, inde facile resistentiam, quam corpus a fluvio sustinet, definire possemus. Tametsi autem ista celeritatis cognitio per rivulum corpori proximum non minoribus difficultatibus sit subiecta, quam determinatio pressionis generatim considerata, tamen hoc inde lucri nanciscimur, ut si nobis licuerit, sive per experientiam, sive undecunque, celeritatem fluidi iuxta corpus praeterlabentis cognoscere, hoc solum nobis satis sit futurum ad veram resistentiam corporis accurate determinandam.*

**"VII. Although the determination of pressure in general—that is, in all points of the fluid—depends both on the paths of individual streams and on the velocity of the water, I have found that if the question is restricted to a single stream, then the pressure at each of its locations can be defined solely by its velocity. Therefore, when the body AME is touched by a single stream f, f, f, and all resistance arises solely from its pressures, if we could know the pressure of this stream at each of its points, we could easily define the resistance that the body sustains from the flow. Although the determination of this velocity of the stream near the body is subject to no fewer difficulties than the general determination of pressure, we nevertheless gain this advantage: if we are able, either through experiment or by some other means, to determine the velocity of the fluid flowing past the body, this alone would suffice for us to accurately determine the true resistance of the body."**

In this passage, the author discusses the complexities of determining the pressure exerted by a fluid on a body, particularly when considering the fluid's velocity along a single stream. The author notes that if the problem is simplified to focus on a single stream, the pressure at various points can be defined solely by the stream's velocity. This insight suggests that if the velocity of the fluid near the body can be determined—whether through experiment or other methods—this would be sufficient to accurately calculate the body's resistance. Despite the challenges associated with measuring this velocity, the author sees it as a potentially fruitful approach to solving the problem of fluid resistance.

*VIII. Si enim ponamus celeritatem, qua aqua circa elementum corporis M praeterlabitur, debitam esse alitudini v, atque assumamus, ut vulgo fieri solet, omnes rivulos in plano horizontali versari, ex iis, quae demonstravi de motu fluidorum in*

*genere, colligitur, pressionem aquae in punto M exprimi per altitudinem $k - v$, ita ut quantitas k pro toto rivulo f, f, f, f, eundem obtineat valorem, ideoque in praesenti negotio pro constanti haberi queat, etiamsi pro diversis rivulis diversos sortiatur valores. Hanc autem presssionem $k - v$ ita interpretari oportet, ut corpus in M a pondere columnae aquae, cuius altitudo sit $= k - v$ sollicitari sit censendum. Pro basi scilicet huius columnae sumi debet elementum superficiei corporis in M, quod ab ista vi normaliter urgebitur, uti in omnibus pressionibus evenit, hincque porro more solito quantitatem totius resistentiae colligere licebit.*

"VIII. For if we assume that the velocity with which the water flows past the element M of the body is proportional to the height $v$, and we assume, as is commonly done, that all streams lie in a horizontal plane, it follows from what I have demonstrated about the motion of fluids in general that the pressure of the water at point M is expressed by the height $k - v$, so that the quantity $k$ holds the same value for the entire stream $f, f, f, f$, and therefore can be considered constant in the present case, even though it may have different values for different streams. This pressure $k - v$ should be interpreted in such a way that the body at M is acted upon by the weight of a column of water whose height equals $k - v$. The base of this column should be taken as the element of the body's surface at M, which will be normally pressed by this force, as occurs in all pressures, and from this we can then proceed in the usual way to calculate the total amount of resistance."

In this passage, the author explains how to calculate the pressure exerted by water on a body at a specific point M. The velocity of the water at M is assumed to be related to a certain height $v$, and the pressure at M is given by the difference $k - v$, where $k$ is a constant for the entire stream. This pressure is understood as the force exerted by a column of water with height $k - v$ on the body's surface at M. The author then suggests using this pressure to calculate the total resistance that the body experiences from the fluid, following standard procedures.

*IX. Quanquam autem circa celeritatem aquae apud singula puncta M nihil habemus exploratum ex Theoria, tamen si experientiam in subsidium vocemus, egregias resistentiae proprietates cognoscemus. Cum enim aucta celeritate in eodem rivulo pressio diminuatur, contra vero augeatur celeritate imminuta, certo affirmare poterimus, in quibus locis corporis AME aqua velocius praeterlabatur, ibi resistentiam esse minorem, quam iis locis, ubi tardius praeterfluit: quae veritas si probe perpendatur, plura alia insignia confectaria suppeditare poterit. Ac merito hoc ingens paradoxon videri debet, quod a maiori celeritate resistentia minor, a minori autem celeritate resistentia maior oriatur; quod primo intuitu regulae vulgari directe adversari videtur. Sed omnis difficultas evanescet, si perpendamus, hic diuersis fluidi celeritates; quibus eodem tempore superficiem corporis stringit, inter se comparari. Neque minus certum manet, si vel fluvius velocius moveatur, vel corpus celerius adversus aquam trudatur, resistentiam quoque maiorem esse futuram.*

"IX. Although we have no theoretical certainty about the velocity of the water at each point M, if we call upon experience for assistance, we will discover significant properties of resistance. For as the velocity in the same stream increases, the pressure decreases, and conversely, as the velocity decreases, the pressure increases. We can therefore confidently assert that where the water flows more quickly past certain parts of the body AME, the resistance will be less than in places where the water flows more slowly. If this truth is carefully considered, it can lead to several other remarkable conclusions. And this should rightly be seen as a great paradox: that greater velocity results in lesser resistance, and lesser velocity results in greater resistance, which at first glance seems to directly contradict the common rule. But all difficulty will disappear if we consider that here, different velocities of the fluid are being compared, which simultaneously interact with the surface of the body. It remains equally certain, however, that if the river flows more quickly, or if the body is pushed more rapidly against the water, the resistance will indeed be greater."

This passage discusses the relationship between fluid velocity and resistance. The author points out that, contrary to what might be expected, increased velocity in a stream results in decreased pressure, and thus decreased resistance, while decreased velocity results in increased pressure and resistance. This paradox challenges the common understanding but can be explained by recognizing that different fluid velocities interact with the surface of the body simultaneously. Despite this, it remains true that if the overall flow of the river is faster or the body moves more quickly against the water, the resistance will increase.

*X. Vicissim ergo ubi per experientiam resistentia maior deprehenditur, ibi celeritas fluidi praeterlabentis minor sit necesse est; cum igitur noverimus, in iis superficiei corporis partibus, ad quas directio fluminis OV propius ad perpendicularem accedit, resistentiam esse maiorem, atque omnium maximam, ubi directio fluvii OV ad corporis superficiem sit normalis; in istis locis quoque celeritas fluidi praeterlabentis minor effe debet. Ad angulum scilicet AMI erit respiciendum, qui quo fuerit maior, sev recto propior, ibi celeritas aquae tanto minor sit necesse est, contra autem eo maior, ubi hic angulus diminuitur. In figura igitur exhibita celeritas aquae praeterlabentis circa A erit minima, circa E vero maxima: atque hoc etiam experientia manifesto declarat, qua constat, aquam circa verticem A plerumque fere penitus stagnare, imprimis si angulus OAM fuerit rectus.*

"X. Conversely, where greater resistance is observed through experience, the velocity of the fluid flowing past must be lower. Therefore, when we know that in those parts of the body's surface where the direction of the stream OV approaches perpendicularity, the resistance is greater—and indeed, it is greatest where the direction of the stream OV is normal to the surface—then the velocity of the fluid flowing past must also be lower in these places. Specifically, attention must be given to the angle AMI, which, the larger it is, or the closer it is to a right angle, the lower the velocity of the water must be in that area. Conversely, the velocity will be greater where this angle is smaller. In the figure provided, the velocity of the water flowing past near point A will be minimal, while near point E it will be maximal. This is also clearly demonstrated by experience, which shows that the water around point A often nearly stagnates, especially if the angle OAM is a right angle."

In this passage, the author explains the relationship between resistance and the velocity of fluid as it flows past different parts of a body. The resistance is greater where the fluid's direction is more perpendicular to the surface, leading to lower fluid velocity in those regions. The angle AMI plays a critical role, with larger angles resulting in slower fluid velocities and smaller angles leading to faster flow. The passage also notes that this concept is confirmed by observation, particularly in areas where water tends to stagnate, such as near point A when the angle OAM is close to a right angle.

*XI. Quoniam igitur novimus per regulam vulgarem, quantumuis debili nitatur fundamento, resistentiam tamen parum a vero aberrantem obtineri, eius beneficio celeritatem aquae iuxta corpus praeterlabentem vero proxime assignare poterimus; et quoniam in eodem rivulo O in singulis locis f f f f E amplitudo reciprocam tenet rationem celeritatis, hinc simul amplitudinem istius rivuli corpus contingentis in singulis locis definire licebit. Tum vero porro primo hoc rivulo constituto simili ratione rivulus sequens f g g f, sev secundus, ex hocque tertius g h h g, indeque sequentes vero proxime designari poterunt. Quae determinationes etsi a veritate aliquantum recedere sunt cesendae, tamen in tam ardua investigatione insigni usu non carebunt. Quodsi enim iam vero proxime tractum singulorum rivulorum una cum aquae celeritate cognoverimus, nullum est dubium, quin deinceps multo facilius summas difficultates, quibus haec quaestio est involuta, superare valeamus. Inde saltem colligere licebit, quemadmodum aequatio generalis figuram singulorum rivulorum complectens debeat esse comparata.*

**"XI. Since we know from the common rule—however weak its foundation may be—that the resistance obtained is not far from the truth, we can use it to approximate the velocity of the water flowing past the body with some degree of accuracy. And since in the same stream $O$, the amplitude at each point $ffffE$ is inversely proportional to the velocity, we can also define the amplitude of the stream touching the body at each point. Then, having determined this first stream, we can similarly determine the next stream $fggf$ (or the second stream), and from this, the third stream $ghhg$, and so on, each being successively approximated. Although these determinations may deviate somewhat from the truth, they will still be of significant use in such a difficult investigation. For if we can approximate the paths of individual streams and the velocity of the water with reasonable accuracy, there is no doubt that we will be much better equipped to overcome the significant difficulties that this question entails. From this, at least, we will be able to deduce how the general equation should be constructed to encompass the shape of each stream."**

In this passage, the author discusses how the common rule, despite its imperfect foundation, can be used to approximate the velocity of water as it flows past a body. By recognizing that the amplitude of the stream is inversely proportional to the velocity, the amplitude of the water touching the body can be determined at each point. This allows for the successive approximation of subsequent streams. Even though these calculations may not be entirely accurate, they can still be valuable in addressing the complex challenges of fluid resistance. The author concludes that once the paths and velocities of individual streams are better understood, it will be easier to formulate a general equation that accounts for the shape of each stream.

*XII. Quodsi autem celeritatem fluvii, qua in notibili a corpore distantia circa O secundum directionem OV movetur, vel, quod eodem redit, celeritatem, qua ipsum corpus AME in aqua stagnante secundum directionem AO fertur, debitam esse ponamus altitudini c, per regulam vulgarem novimus, ubi corporis superficies ad diretionem fluminis sit perpendicularis, ibi resistentiam exprimi per ipsam altitudinem c, fia autem in loco M angulus incidentiae AMI, ducta recta MI directioni AO parallela, ponatur* $= \phi$, *per eandem regulam constat, fore resistentiam in* $M = c\sin^2\phi$. *Hinc ergo, comparatione instituta, si aquae iuxta corpus praeterfluentis celeritas in M debita statuatur altitudini* $v$, *hanc adipiscemur aequationem:*

$$c\sin^2\phi = k - v, \text{ ideoque } v = k - c\sin^2\phi$$

*Quocirca ex hac formula veram aquae celeritatem ad singula corporis puncta M assiguare valebimus.*

"XII. However, if we assume that the velocity of the stream at a noticeable distance from the body, moving around point $O$ in the direction $OV$, or, equivalently, the velocity at which the body $AME$ moves in stagnant water in the direction $AO$, is proportional to the height $c$, we know from the common rule that where the surface of the body is perpendicular to the direction of the stream, the resistance is expressed by the height $c$. But if at point $M$, the angle of incidence $AMI$, formed by drawing a straight line $MI$ parallel to the direction $AO$, is denoted as $\phi$, then by the same rule, the resistance at $M$ is given by $c\sin^2\phi$. Therefore, by comparison, if the velocity of the water flowing past the body at point $M$ is proportional to the height $v$, we obtain the equation:

$$c\sin^2\phi = k - v,$$

and consequently,

$$v = k - c\sin^2\phi.$$

Thus, from this formula, we will be able to determine the true velocity of the water at each point $M$ on the body."

In this passage, the author explains how to calculate the true velocity of water at specific points on a body submerged in a stream. Starting with the assumption that the velocity of the stream far from the body is proportional to a certain height $c$, and using the common rule that expresses resistance when the surface is perpendicular to the flow, the author derives an equation that relates the velocity at any point $M$ on the body to the angle of incidence $\phi$. This equation allows for the calculation of the actual velocity of the water at each point on the body's surface.

*XIII. Tantum ergo superest, ut hinc constantem quantitatem k definiamus, quae quidem ex casu, ubi angulus* $\phi$ *est rectus, facile colligetur. Experientia enim testatur, in his locis celeritatem aquae allabentis esse nullam, tum vero etiam nulla adest ratio, cur aqua, ubi directio fluminis ad superficiem corporis est perpendicularis, in hauc potius plagam, quam aliam, dilaberetur. Ex quo conficitur, si angulus* $\phi$ *fuerit rectus, ideoqne* $\sin\phi = 1$, *tum este oportere* $v = 0$; *unde manifesto sit* $k = c$, *sev ista constans k praecise est aequalis alitudini fluminis celeritati debitae. Posita autem* $k = c$, *habebimus,* $v = c - c\sin^2\phi$, *sev* $v = c\cos^2\phi$, *hincque* $\sqrt{v} = \cos\phi \cdot \sqrt{c}$:

*unde hanc insignem proprietatem derivamos, quod celeritas aquae iuxta corpus ad M praeterlabentis sit ad veram celeritatem fluminis $\sqrt{c}$, uti cosinus anguli AMI ad sinum totum. Atque hinc in E, ubi tangens directioni OA est parallela, sev $\phi = 0$ erit $v = c$ , sev celeritas aquae ibi aequalis resultabit ipsi fluminis celeritati in O.*

"XIII. Thus, it only remains to define the constant quantity $k$, which can easily be determined from the case where the angle $\phi$ is a right angle. For experience shows that in these locations, the velocity of the water flowing by is zero, and there is no reason why water, when the direction of the stream is perpendicular to the surface of the body, should flow in one direction rather than another. From this, it follows that if the angle $\phi$ is a right angle, and hence $\sin\phi = 1$, then $v$ must be zero. Therefore, it is clear that $k = c$, or that this constant $k$ is precisely equal to the height corresponding to the velocity of the stream. Thus, with $k = c$, we obtain:

$$v = c - c\sin^2\phi, \quad \text{or} \quad v = c\cos^2\phi.$$

Hence,

$$\sqrt{v} = \cos\phi \cdot \sqrt{c}.$$

From this, we derive the remarkable property that the velocity of the water flowing past the body at point $M$ is to the true velocity of the stream $\sqrt{c}$ as the cosine of the angle $AMI$ is to the sine of the whole angle. Consequently, at point $E$, where the tangent is parallel to the direction $OA$, or $\phi = 0$, $v = c$, meaning that the velocity of the water there will be equal to the velocity of the stream at point $O$."

In this passage, the author derives the relationship between the velocity of water flowing past a body and the angle of incidence. By setting the constant $k$ equal to the height corresponding to the stream's velocity, the author finds that the velocity $v$ at any point $M$ on the body can be expressed in terms of the cosine of the angle $\phi$. This leads to the conclusion that the velocity of the water flowing past the body is proportional to the cosine of the angle, with the maximum velocity occurring when the tangent is parallel to the direction of the stream.

*XIV. Hinc discimus, si celeritatem navis, qua vehimur, ex velocitate aquae praeterlabentis aestimare velimus, atque navis secundum directionem OA progrediatur, tum in navi eum locum E esse eligendum, ubi tangens horizontalis directioni AO sit parellala. Atque in hoc loco tuto concludere poterimus, celeritatem navis aequalem esse velocititi aquae, quae hic praeterlabitur: sin autem in alio loco, uti in M, hoc indicium instituere vellemus, eo magis erraremus, quo maior fuerit angulus AMI, navem scilicet nimis parvam reputantes; quoniam celeritas aquae in M praeterlabentis minor est celeritate navis, et quidem in ratione cofinus anguli AMI ad sinum totum. Interim tamen probe est recordandum, has determinationes non summo rigore esse veras, sed tantum idoneas ad veritatem appropinquationes.*

"XIV. From this, we learn that if we wish to estimate the speed of a ship by the velocity of the water flowing past it, and the ship is moving in the direction $OA$, then the point $E$ on the ship, where the horizontal tangent is parallel to the direction $AO$, should be chosen. At this point, we can safely conclude that the speed of the ship is equal to the velocity of the water flowing past. However, if we were to make this determination at another point, such as at $M$, we would err more significantly the larger the angle $AMI$ is, underestimating the speed of the ship, because the velocity of the water flowing past at $M$ is less than the speed of the ship, and indeed in the ratio of the cosine of the angle $AMI$ to the whole sine. However, it should be well remembered that these determinations are not strictly accurate, but are merely suitable approximations to the truth."

In this passage, the author discusses how to estimate the speed of a ship by observing the velocity of the water flowing past it. The key point is that the best location on the ship for such an observation is where the tangent to the ship's surface is parallel to the ship's direction of motion. At this point, the water's velocity matches the ship's speed. Observing at other points, especially where the angle between the direction of motion and the surface is large, will lead to underestimating the ship's speed due to the slower water flow at those points. The author emphasizes that these determinations are approximations, not exact values.

*XV. Ac regula quidem haec certo fallit in corporis parte posteriori ENB; si enim ponamus, ut in parte anteriori, esse $v = c\,cos^2\,\phi$, puppis navis praecise tanta vi propelleretur, quanta prora repellitur; unde a puncto E retrorsum formula $v = c\,cos^2\,\phi$ eo magis a veritate discedet, quo propius ad B perveniamus; tantum ergo in parte anteriori AME, tanquam toleranter vera, admitti potest. Interim tamen hinc coniectando suspicari poterimus, quomodo motus aquae praeterlabentis circa puppim navis ENB se sit habiturus. Si enim puppis nihil ad resistentiam conferat; certum est, aquam ab E ad B celeritate uniformi defluere, ea scilicet, quae debeatur altitudini c, et quam iam in E recuperavit. Sin autem in hac parte lentius decurrat, navis hinc propulsionem accipiet, qua resistentia diminuetur. Fieri autem nequit, ut usquam evadat v > c, quia tunc pressio prodiret negativa. Hoc enim casu acqua post navim vacuum relinqueret, et navis quasi sulcum traheret; unde ob deficientem pressionem a tergo resistentia utique augeretur.*

Fig. 2.

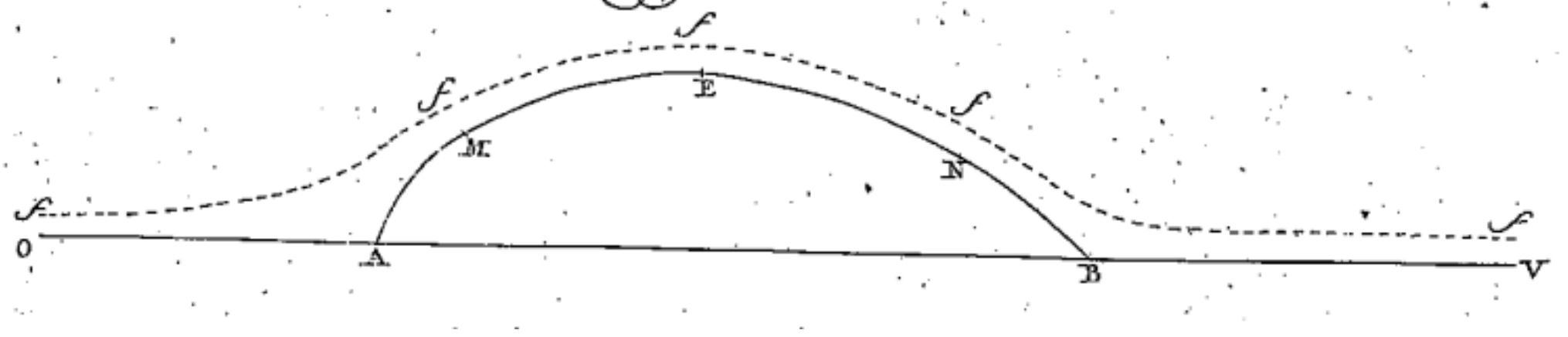

"XV. This rule certainly fails at the rear part of the body $ENB$. If we assume, as in the front part, that $v = c \cdot \cos^2 \phi$, the stern of the ship would be propelled with precisely the same force as the bow is repelled. Thus, from point $E$ backwards, the formula $v = c \cdot \cos^2 \phi$ will deviate further from the truth the closer we get to $B$. Therefore, it can only be accepted as approximately true in the front part $AME$. However, from this, we may conjecture and speculate about the motion of the water flowing past the stern of the ship $ENB$. If the stern contributes nothing to the resistance, it is certain that the water will flow uniformly from $E$ to $B$ with the velocity $c$, which it regained at $E$. But if the flow is slower in this part, the ship will receive propulsion from it, reducing the resistance. However, it is impossible for $v$ to exceed $c$ at any point because that would result in negative pressure. In such a case, the water would leave a vacuum behind the ship, and the ship would drag a furrow, increasing resistance due to the lack of pressure from behind."

In this passage, the author discusses the limitations of the previously established rule for determining the speed of water flowing past a ship. Specifically, the rule does not hold true at the rear part of the ship, where the water's velocity may not behave as predicted. The author explains that if the water flows slower at the stern, the ship may experience propulsion, reducing resistance. However, if the velocity exceeds a certain threshold, negative pressure could occur, causing a vacuum behind the ship and increasing resistance. This highlights the complexities of fluid dynamics around a moving ship.

*XVI. Si igitur puppi navis ENB eiusmodi figura tribui posset, ut aqua ab E et B progrediendo retardaretur; atque circa N et B minorem habitura esset celeritatem, quam in E, talis figura conftructioni navium esset aptissima iudicanda, quia hoc modo aqua puppim adeo antrorsum propelleret, resistentiamque prorae diminueret. Verum si experientiam consulamus, talem figuram vix dari colligere licet, quin potius omnis cura eo conferri debere videtur, ut ne alterum incommodum usu veniat, quo ob vacuum pene navem relictum resistentia adeo augetur. In eo imprimis ergo circa figuram puppis erit elaborandum, ut tale vacuum evitetur, ac puppis ita insensibiliter ad B usque convergat, ut aqua eam iugiter sequatur, neque rivulus Ef eam usquam deferat. In hoc etiam insignis illa navium proprietas versatur, qua puppi talem figuram conciliare student, ut aqua libere ad gubernaculum decurrere queat, quoeffectu frustraremur, si aqua circa puppim navem defereret, neque in gubernaculum allideret.*

**XVI.** If the stern of the ship $ENB$ could be given such a shape that the water is slowed as it progresses from $E$ to $B$, and has a lower velocity around $N$ and $B$ than at $E$, such a design would be considered optimal for ship construction. In this way, the water would push the stern forward, reducing the resistance at the bow. However, if we consult experience, it seems unlikely that such a shape exists. Rather, all efforts should focus on avoiding the other issue, where resistance significantly increases due to the near-vacuum left behind the ship. Therefore, particular attention must be paid to the shape of the stern to prevent such a vacuum, ensuring that the stern gradually converges to point $B$ in a way that the water continuously follows it, and that the current $Ef$ does not detach from it. This also relates to the notable property of ships, where they aim to shape the stern so that water can flow freely to the rudder, which would be impeded if the water were to detach from the stern and not strike the rudder.

This section of the text discusses the design considerations for the stern of a ship. The author suggests that if the stern could be shaped in such a way that water slows down as it flows towards the rear, it would reduce resistance and improve propulsion. However, practical experience indicates that achieving such a shape may be difficult. Instead, the focus should be on preventing the formation of a vacuum behind the ship, which would increase resistance. The stern should be designed to gradually taper, ensuring that water follows the ship closely and flows smoothly to the rudder, maintaining control and reducing drag.

*XVII. Ex celeritate autem aquae iuxta corpus defluentis figuram rivulorum illorum, per quos aqua motum suum inflectit; satis exacte colligere poterimus. Ac primo quidem pro rivulo corpori proximo fff eius amplitudo ubique celeritati reciproce debet esse proportionalis. Cum igitur, posito angulo AMI =* $\phi$*, celeritas aquae in M sit = c cos* $\phi$*, in hoc loco amplitudo rivuli f erit ut* $\frac{1}{c \cdot \cos\phi}$*; quia autem hunc rivulum angustissimum concipimus, motusque aquae in M secundum curvae tangentem dirigitur, amplitudo Mq ad curvam statuenda est normalis. Quare in normali QM producta capiatur portio Mq, quae sit ubique ut* $\frac{1}{c \cdot \cos\phi}$*, sev ut sec* $\phi$*, ob c constantem, et punctum q erit in curva proxima fgqe rivulum exhibente. Verum hic* $\phi$ *denotabit quoque angulum PMQ, posita applicata PM ad fluminis directionem OA perpendiculari: unde erit Mq ut* $\frac{MQ}{MP}$*. Producatur ergo ubique applicata PM in p, ut pars producta Mp sit constantis magnitudinis, et ex p axi AO agatur parallela pq, normalem QM productum secans in q, eritque punctum q in curva quaesita.*

Fig. 3.

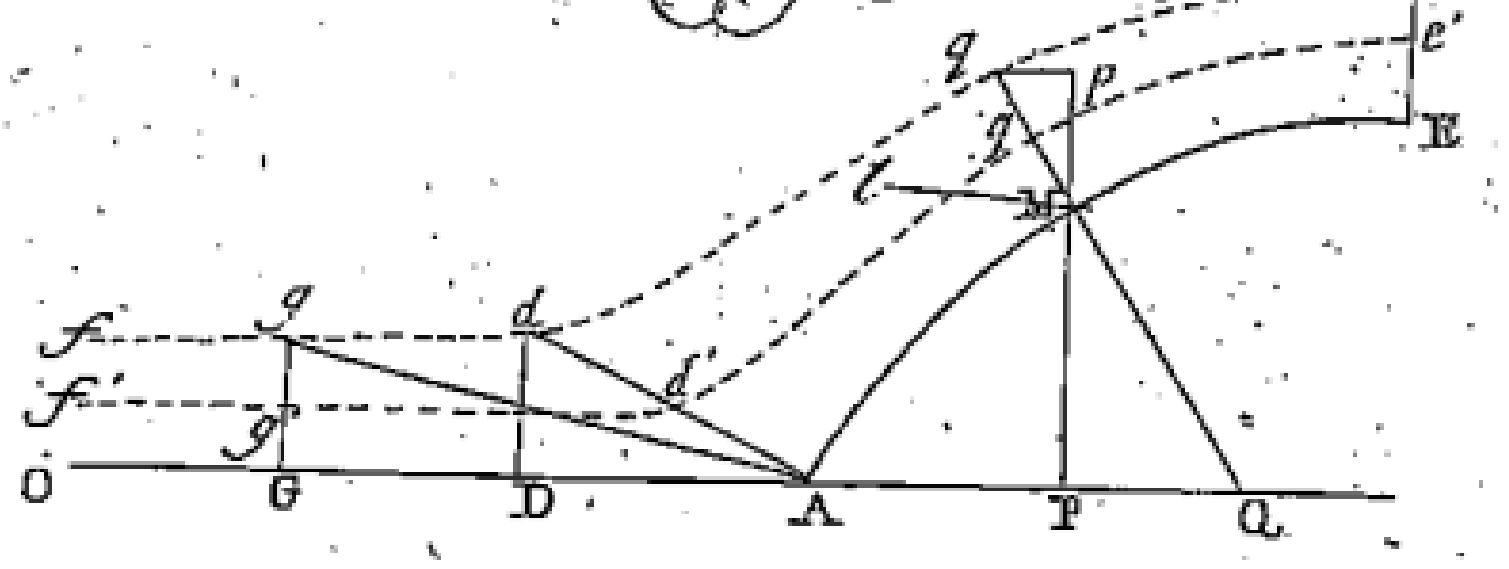

**XVII.** From the velocity of water flowing along the body, we can deduce the shape of those streams through which the water alters its path with reasonable accuracy. First, for the stream closest to the body $fff$, its width should be inversely proportional to the velocity at every point. Therefore, if we assume the angle $AMI = \phi$, and the velocity of water at $M$ is $c \cdot \cos\phi$, then at this point, the width of the stream $f$ will be proportional to $\frac{1}{c \cdot \cos\phi}$. Since we conceive of this stream as being the narrowest and the water's motion at $M$ is directed along the tangent of the curve, the width $Mq$ should be measured perpendicular to the curve.

Therefore, on the normal $QM$ extended, the portion $Mq$ should be taken to be proportional to $\frac{1}{c \cdot \cos\phi}$, or simply proportional to $\sec\phi$, given that $c$ is constant, and the point $q$ will lie on the adjacent curve $fgqe$, which represents the stream. Here, $\phi$ also denotes the angle $PMQ$, with the ordinate $PM$ perpendicular to the flow direction $OA$. Thus, $Mq$ will be proportional to $\frac{MQ}{MP}$.

Now, extend each ordinate $PM$ to $p$, such that the length $Mp$ is of a constant magnitude. Then, draw a parallel line $pq$ to the axis $AO$, intersecting the normal $QM$ at $q$, and the point $q$ will lie on the desired curve.

This passage explores how the shape of the streams (or rivulets) near a ship's body can be derived from the velocity of the water as it flows past. By understanding that the stream width is inversely proportional to the water's velocity at each point, the author provides a method to determine the curvature of the streams. The mathematical relationships between the velocity, angles, and distances help construct the curve that represents the flow pattern around the ship's body.

*XVIII. Cum ergo curva fgqe hac praedita sit proprietate, ut sit intervallum Mp constantis magnitudinis, in puncto E, ubi tangens curvae est axi AO parallela, ipsa rivuli amplitudo Ee, quae est applicatae PM parallela, hanc amplitudinem habebit, sev vicissim intervallum Mp ubique isti amplitudini Ee aequale est capiendum, unde patet, quemadmodum ab E per M ad A progrediendo amplitudo rivuli continuo augeatur. Hinc ergo pro vertice corporis A, si recta Ad fuerit ad curvam normalis, puncti d ab axe AO distantia Dd quoque intervallo Ee erit aequalis, et quoniam hic rivuli amplitudo per ipsam rectam Dd aestimari debet, in hoc loco Dd aquae celeritas aequalis est censenda celeritati in Ee, hoc est verae fluminis celeritati, ita ut hic fluvius adhuc vero suo motu feratur, neque ullam ob corpus oppositum mutationem fubierit. Quin etiam, si corpus in A angulo terminetur, quaelibet alia recta Ag ad axem AO magis inclinata pariter ad curvam in A normalis est censenda, unde et hic distantia ab axe Gg ipsi Ee et Dd est aequalis, sicque ultra d rivulus includetur recta dgf, axi AO parallela.*

**XVIII.** Given that the curve $fgqe$ has the property that the interval $Mp$ is of constant magnitude, at point $E$, where the tangent of the curve is parallel to the axis $AO$, the width of the stream $Ee$ (which is parallel to the ordinate $PM$) will have the same width as this interval. Conversely, the interval $Mp$ at every point should be considered equal to the width $Ee$, which shows how, as one progresses from $E$ through $M$ to $A$, the width of the stream continuously increases.

Therefore, at the vertex of the body $A$, if the line $Ad$ is normal to the curve, the distance $Dd$ of point $d$ from the axis $AO$ will also be equal to the interval $Ee$. Since the stream width is to be estimated by this line $Dd$, at this point, the water velocity is to be considered equal to the velocity at $Ee$, which is the true velocity of the stream. Thus, the stream still moves with its true motion and has not undergone any change due to the body being in its path.

Moreover, if the body terminates at $A$ with an angle, any other line $Ag$ more inclined to the axis $AO$ will also be considered normal to the curve at $A$, and hence the distance from the axis $Gg$ will be equal to both $Ee$ and $Dd$. Therefore, beyond point $d$, the stream will be enclosed by the line $dgf$, which is parallel to the axis $AO$.

This section provides a geometric analysis of how the width of the stream changes as it flows around the body of the ship, from point $E$ to point $A$. The key takeaway is that the width of the stream increases as it progresses along the body, and at certain key points, the velocity of the water remains unchanged, indicating that the stream continues with its natural motion despite the presence of the body in its path. This exploration of stream behavior near the body has implications for understanding fluid dynamics around objects in motion.

*XIX. En ergo figuram primi rivuli fgdqe corpori AME proximi et altera parte cum axe AO tum corpore AME terminati, per cuius partem anteriorem OfDd aqua motu suo naturali affluit. Cum autem ultra Dd ad corpus appropinquaverit, ob crescentem amplitudinem rivuli, eius motus partim retardabitu, partim diectionem ita inflectet, ut ab f ad q usque directionem quidem curvae fq sequatur, ex altera vero parte primum secundum axem DA, tum vero secundum ductum curvae AM progrediatur; atque ad A ob maximam amplitudinem motu minimo feratur. At vero singula intervalla Ee, Mq, Dd, Gg infinite parva sunt concipienda, quae si denuo in duos pluresue rivulos minores subdividentur, uti in figura bisectio per lineam f'g'd'q'e' repraesentatur, unde motus aquae per singulos hos rivulos eiusque retardatio et inflexio multo clarius perspicitur.*

**XIX.** Behold the figure of the first stream $fgdqe$ adjacent to the body $AME$, bordered on one side by the axis $AO$ and on the other by the body $AME$. Along the front part of the body, the water flows naturally from $OfDd$. However, as it approaches the body beyond point $Dd$, due to the increasing width of the stream, its motion will be partly slowed and partly redirected, such that from $f$ to $q$, the flow will follow the curve $fq$, and from that point, it will progress first along the axis $DA$, and then along the curve $AM$. At point $A$, due to the maximum width of the stream, the motion will be minimal.

Moreover, all the intervals $Ee$, $Mq$, $Dd$, and $Gg$ should be considered infinitesimally small. If these intervals are further subdivided into two or more smaller streams, as depicted by the bisection through the line $f'g'd'q'e'$ in the figure, the motion of the water through these smaller streams, as well as its deceleration and redirection, will become even clearer.

This passage emphasizes how the stream's width and flow characteristics change as the water moves around the body $AME$. The increasing width of the stream slows down the water and causes it to change direction. The concept of subdividing the stream into smaller parts helps in understanding the nuanced behavior of water flow and how it adjusts around obstacles. This section of the text is particularly relevant to fluid dynamics, as it sheds light on the complex interactions between fluid flow and solid bodies.

*XX. Quanquam haec tantum proxime ad veritatem accedere sunt censenda, atque adeo ultra A versus O lex continuitatis in formula nostra non amplius observatur, cum vi formulae amplitudo rivuli in d non per rectam Dd sed Ad esset aestimanda, tamen haec ita ad veritatem, quam experientia monstrare solet, accedere videntur, ut si non per hanc ipsam constuctionem, tamen per satis similem vera figura singulorum rivulorum definiri sit censenda. Per experientiam enim certum est, tantum in modica a corpore distantia motum demum fluminis perturbari incipere, ita ut, cum retardetur, tum circa corpus inflectatur, omnino uti delineatio rivulorum secundum formulam nostram facta manifesto declarat. Atque in parte corporis antica AME nullum est dubium, quin intervalla lateralia Mp sint inter se proxime aequalia, pone corpus autem, ut vidimus, haec aequalitas cessabit, dum ibi ipsae amplitudines Mq potius aequalitatis legem sequi videntur.*

**XX.** Although these approximations are considered to be close to the truth, the law of continuity in our formula is no longer observed beyond point $A$ towards $O$. By the formula, the width of the stream at $d$ should be estimated not by the line $Dd$, but by $Ad$. However, these results seem to approach the truth as shown by experience, suggesting that if the actual figure of the streams cannot be determined by this precise construction, it can be closely estimated by a similar one.

It is certain through experience that only at a short distance from the body does the flow begin to be disturbed. As the flow slows down and bends around the body, this behavior aligns with the depiction of the streams made according to our formula. In the front part of the body $AME$, there is no doubt that the lateral intervals $Mp$ are approximately equal to each other. However, behind the body, as we have seen, this equality ceases, and there the widths $Mq$ appear to follow the law of equality instead.

This passage further refines the analysis of fluid flow around a body, emphasizing that while the formula provides a good approximation, the actual behavior of the fluid may deviate from it, especially beyond certain points. The mention of the law of continuity not being maintained highlights the complexities involved in fluid dynamics, particularly in the regions where the flow is disturbed. The idea that the formula is validated by experience underlines the importance of empirical observations in confirming theoretical models.

*XXI. Ut a simplicioribus incipiam, terminetur corporis pars antica duabus lineis rectis AE et EF, quaquarum haec sit directioni fluminis parallela, illa utcunque inclinata; haec scilicet figura quasi semissis corporis est spectanda, iudiciumque partis ultra rectam AC sitae pari modo absolvetur, dummodo punctum A maxime promineat. Iam ad rivulos designandos ad rectam inclinatam AE ducantur normales Ad, Ee, tum in dato intervallo Dd = Ff, directioni fluminis OA parallelae agantur od, fe, iunganturque puncta d et e recta de; ac linea composita odef repraesentabit tractum rivuli proximi, simili vero modo si intervalla D'd', F'f' maiora capiantur, figura rivuli sequentis o'd' e'f' prodibit. Sic quidem secundum regulam inventam figura rivulorum exprimetur; revera autem circa d et e anguli obtundentur, quia aqua nom subito, sed successive, directionem mutabit: unde quo magis rivuli a corpore distabunt, eo magis eorum tractus ad uniformitatem accedet, quin etiam intervalla Ff ratione Dd ita insensibiliter diminuentur, ut tandem rivuli satis remoti directioni OA plane paralleli restituantur.*

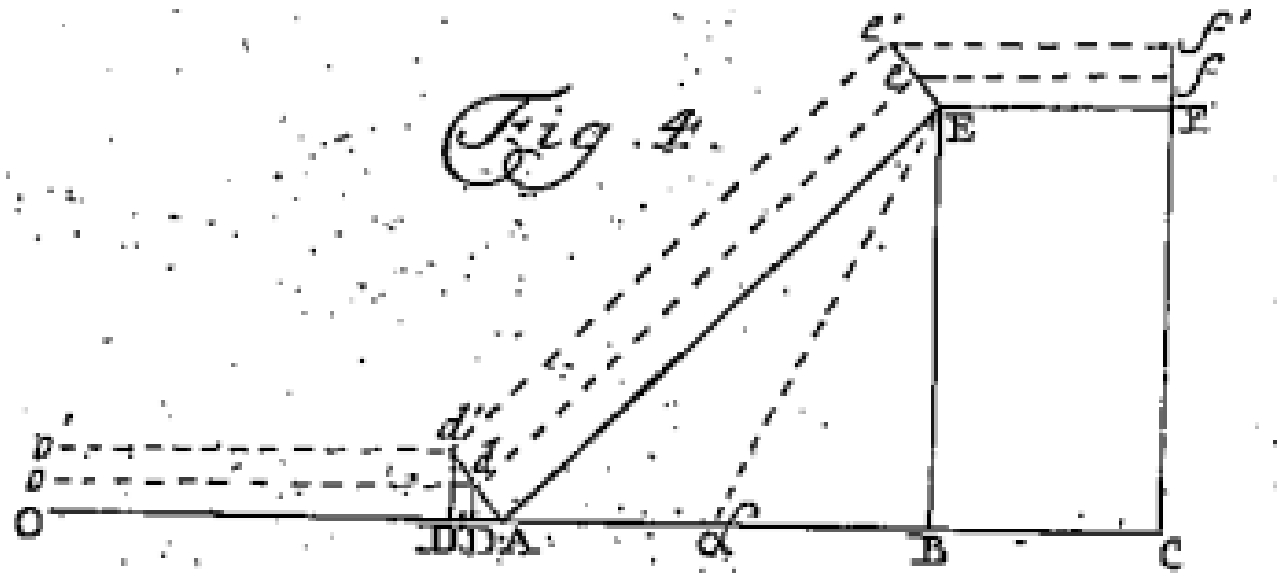

**XXI.** To begin with simpler cases, let the front part of the body be bounded by two straight lines, $AE$ and $EF$, with $EF$ being parallel to the direction of the flow and $AE$ inclined at an arbitrary angle. This shape is essentially half of the body, and the evaluation of the part beyond the line $AC$ will be completed similarly, provided that point $A$ protrudes the most.

To designate the streams, draw perpendiculars $Ad$ and $Ee$ to the inclined line $AE$. Then, in the given interval $Dd = Ff$, draw lines parallel to the flow direction $OA$ from $d$ and $f$, and connect points $d$ and $e$ with a straight line $de$. The composed line $odef$ will represent the path of the closest stream. Similarly, if larger intervals $D'd'$ and $F'f'$ are taken, the path of the next stream $o'd'e'f'$ will be produced.

Thus, according to the found rule, the shape of the streams will be expressed. However, in reality, the angles around $d$ and $e$ will be blunted because the water will change direction not suddenly but gradually. Consequently, the further the streams are from the body, the more their paths will approach uniformity. Furthermore, the intervals $Ff$ will decrease so gradually relative to $Dd$ that eventually, the streams will restore to being parallel to the direction $OA$ once they are sufficiently distant from the body.

This passage explores the behavior of fluid streams around a simple body shape. It explains how to construct the paths of the streams using geometric methods and anticipates the gradual change in the streams' direction as they move away from the body. The observation that distant streams eventually become parallel to the flow direction emphasizes the tendency toward uniformity in fluid motion after the initial disturbance caused by the body's presence.

*XXII. In primo ergo rivulo aqua per totum tractum OoDd celeritatem suam et directionem retinebit, ac mutatio demum in distantia AD a corpore incipiet, nisi quatenus ob incurvationem ad d hoc intervallum aliquantum augeri est censendum. Cum igitur sit Dd:AD = AB:BE, erit ista distantia ante corpus, in qua motus aquae perturbari incipit, $AD = \frac{BE}{AB} \cdot Dd = Dd \cdot \tan BAE$. Unde si angulus BAE fuerit rectus, hoc spatium in infinitum augeri videtur; verum cum ipsa rivuli amplitudo Dd pro infinite parva sit habenda, hinc intervallum ad magnitudinem finitam redigetur. Verum si plures positiones lateris EA, ut Eα, inter se comparemus, quae omnes eadem latitudine BE sint praeditae, ponamusque BE = a, BA = x, et amplitudinem rivuli Dd = Ff = f, locus D, ubi motus aquae primum perturbari incipit, a recta BE distabit intervallo $BD = x + \frac{af}{x}$, quod sit omnium minimum, si $x = \sqrt{af}$, seu $B \propto = \sqrt{BE \cdot Dd}$, quo casu angulus $B \propto E$ iam minime a recto distabit. Verisimile autem est, si spatium $B \propto$ adhuc minus capiatur, atque adeo evanescat, intervallum BD non fieri magis, cum positio BE non in maiori distantia motum aquae perturbare valeat, quam positio $\propto E$, unde et pro positione BE haec distantia erit censenda $BD = 2\sqrt{af}$.*

**XXII.** In the first stream, the water will maintain its speed and direction throughout the entire path $OoDd$, with changes only beginning at the distance $AD$ from the body, unless, due to curvature at $d$, this interval is expected to increase slightly. Since $Dd : AD = AB : BE$, the distance before the body where the water's motion starts to be disturbed will be $AD = \frac{BE}{AB} \times Dd = Dd \times \tan BAE$. If the angle $BAE$ is a right angle, this distance seems to increase infinitely. However, since the stream's width $Dd$ is considered infinitely small, this interval will be reduced to a finite magnitude.

Now, if we compare multiple positions of the side $EA$, such as $E\alpha$, which all have the same width $AE$, and let $BE = a$, $BA = x$, and the stream width $Dd = Ff = f$, then the point $D$, where the water's motion first becomes disturbed, will be at a distance $BD = x + \frac{af}{x}$ from the line $BE$. This distance will be minimal when $x = \sqrt{af}$, making $B \propto = \sqrt{BE \times Dd}$, where the angle $B \propto E$ will be only slightly less than a right angle.

It is likely that if the distance $B \propto$ is reduced further, to the point of vanishing, the interval $BD$ will not increase any more, as the position $BE$ will not disturb the water's motion at a greater distance than the position $\propto E$. Hence, for the position $BE$, this distance should be considered as $BD = 2\sqrt{af}$.

In this passage, the author explains the behavior of water in the first stream near a body. The analysis involves calculating the distance at which the water's motion first becomes disturbed and how this distance changes with different body positions and stream widths. The conclusion suggests that the disturbance starts at a minimal distance from the body, depending on the geometry and dimensions involved.

*XXIII. Hinc ergo colligere poterimus, quomodo aqua ad superficiem BE, quae ad directionem fluminis est normalis, alluat. Scilicet rivulus od, cuius ab axe OB distantia sit Dd, motu inalterato affluet usque ad d, ut sit distantia* $BD = 2\sqrt{BE \cdot Dd}$, *hicque demum motum suum inflectet ad e progrediens, unde secundum ef lateri EF parallele proferetur, ut sit distantia Ff = Dd. Simili modo rivulus remotior viam sequetur o'd'e'f', cursum suum iam in d' inflectens, ut sit intervallum BD'=* $2\sqrt{BE \cdot D'd'}$. *In spatiis autem Bd et dd', quia ibi amplitudo rivulorum est maxima, motus aquae erit tardissimus, et ad B penitus quiescet, unde hic resistentia quoque erit maxima, ad E versus F autem, ob rivuli primi amplitudinem decrescentem, continuo diminuetur, neque tamen diminutio tanta esse potest, ut resistentia inde orta a regula vulgari notabiliter abhorreat. Haud aliter resistentia comparata fore videtur, si latus EB retro fuerit inclinatum.*

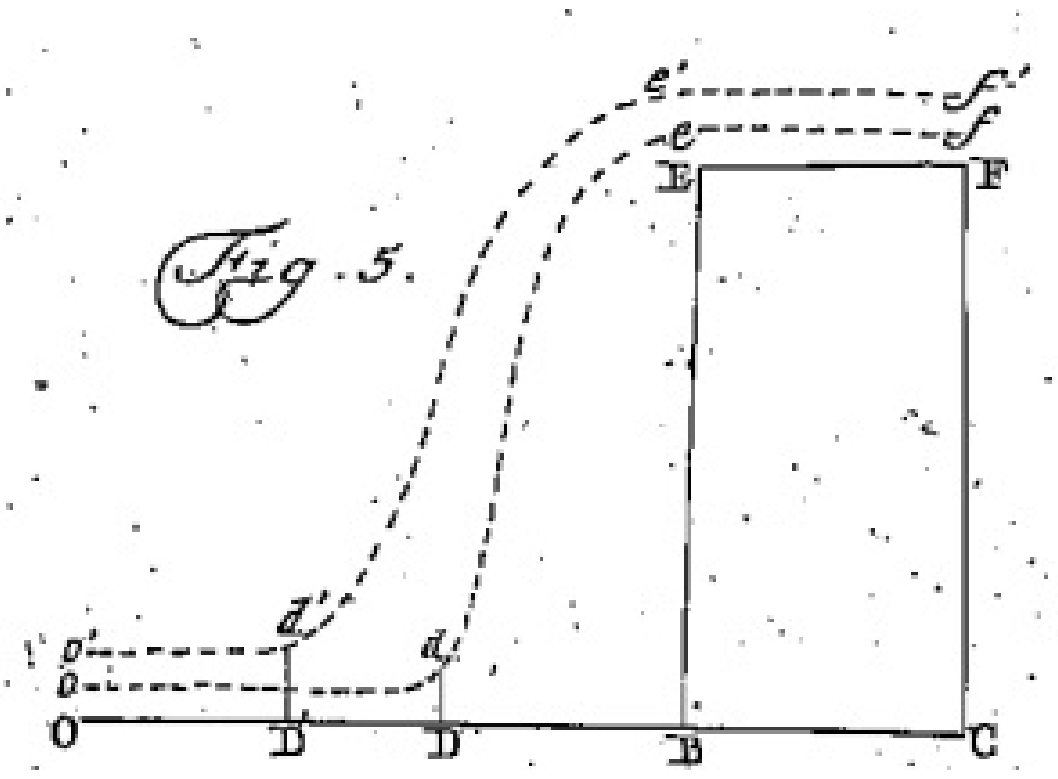

**XXIII.** From this, we can infer how the water flows against the surface $BE$, which is perpendicular to the direction of the current. Specifically, the stream $od$, at a distance $Dd$ from the axis $OB$, will flow undisturbed until it reaches $d$, at which point the distance $BD$ will be $2\sqrt{BE \times Dd}$. Here, the flow will bend towards $e$, continuing parallel to the side $EF$ along $ef$, where the distance $Ff$ equals $Dd$.

Similarly, a more distant stream will follow the path $o'd'e'f'$, changing direction at $d'$, with the interval $BD'$ being $2\sqrt{BE \times Dd'}$. In the regions $Bd$ and $dd'$, where the stream's width is greatest, the water's movement will be slowest and will nearly stop at $B$, leading to maximum resistance. However, towards $E$ and $F$, due to the decreasing width of the first stream, the resistance will gradually decrease. Nevertheless, this reduction in resistance cannot be so significant as to deviate notably from the usual rule.

Similarly, the resistance would likely be comparable if the side $EB$ were inclined backward.

This passage describes the flow of water near the surface $BE$, perpendicular to the current, and explains how the water moves around the body, leading to variations in resistance. The author also discusses how the resistance changes along the surface, particularly as the flow bends and moves parallel to different parts of the body.

*XXIV. Sit iam corporis figura AMF quadrans circuli, atque, ad tractum rivuli proximi inueniendum, ponatur radius circuli CA = CM = a, amplitudo rivuli in F, nempe Ff = f. Pro puncto quocunque circuli M ponatur abcissa CP = x, applicata PM = y, ut sit xx + yy = aa. Tum producto radio CM in m, ut sit applicata curvae quesita pm = y+ f, erit abscissa* $Cp = x + \frac{fx}{y}$*. Statuantur ergo pro curva omf coordinatae Cp = X, pm = Y, ut sit Y = y + f et* $Cp = \frac{(y+f)x}{y} = \frac{Yx}{y}$*; eritque y = Y - f et* $x = \frac{X(Y-f)}{Y}$ *unde ob xx + yy = aa pro curva omf habebitur haec aequatio* $(XX + YY)(Y-f)^2 = aaYY$*: quae si f ut parameter variabilis spectetur, innumerabiles istiusmodi curvas omf exhibebit; quae omnes secundum axem AO in infinitum extendentur, ab eoque tandem intervalo = f distabunt, unico casu excepto, quo f = 0 ipsum circulum AMF referente. Cum enim sit* $XX = \frac{aaYY}{(Y-f)^2} - YY$*, si X in infintum abeat, fiet Y = f. Neque vero omnes hae curvae rivulos exhibebunt, propterea quod quaeque sequens non eodem modo ex praecedente definitur, uti prima ex ipso circulo est constructa.*

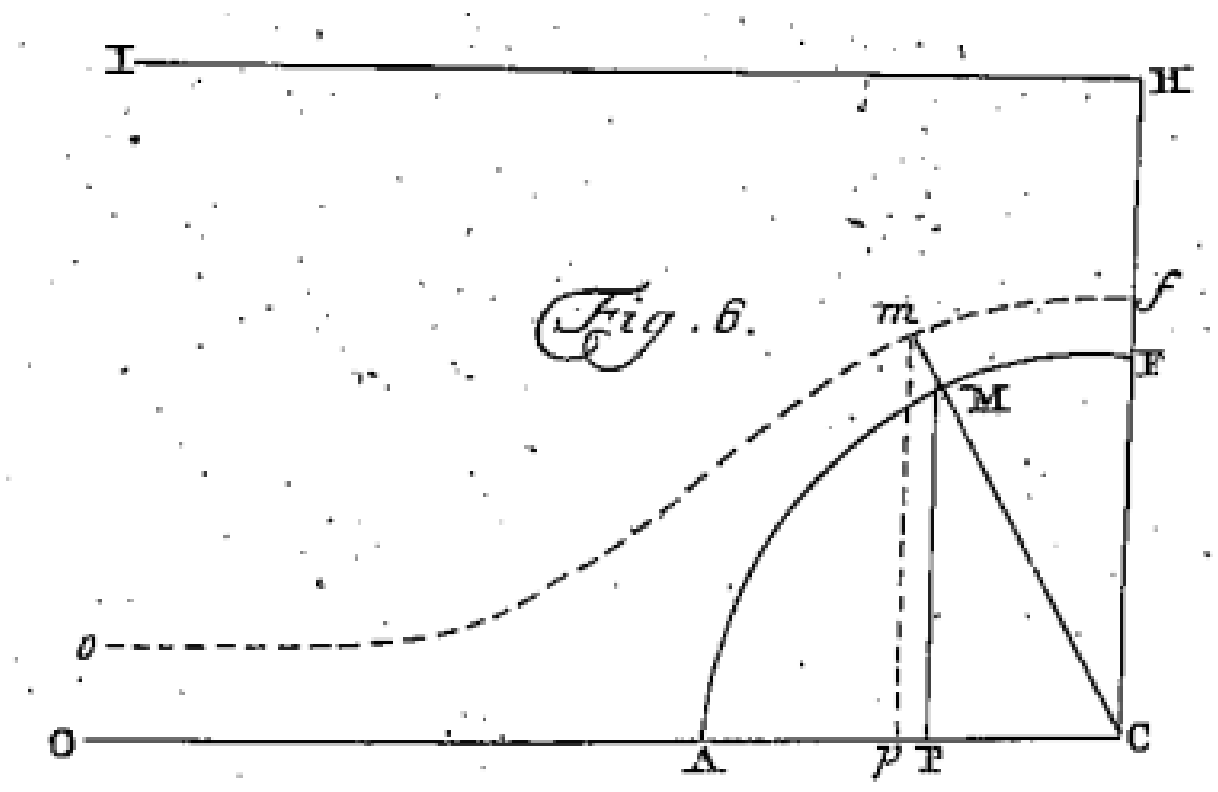

**XXIV.** Now consider the shape of the body $AMF$ as a quarter of a circle. To find the path of the nearest stream, let the radius of the circle $CA = CM = a$, and the width of the stream at $F$ be $Ff = f$. For any point $M$ on the circle, let $CP = x$ be the abscissa, and $PM = y$ the ordinate, so that $x^2 + y^2 = a^2$. Then, extending the radius $CM$ to $m$, such that the ordinate of the desired curve $pm = y + f$, the abscissa $Cp$ becomes $x + \frac{fx}{y}$.

Thus, for the curve $omf$, the coordinates $Cp = X$ and $pm = Y$, yield $Y = y + f$ and $Cp = \frac{(y+f)x}{y} = \frac{Yx}{y}$. Hence, $y = Y - f$ and $x = \frac{X(Y-f)}{Y}$, leading to the equation:

$$(XX + YY) \times (Y - f)^2 = a^2 \times YY$$

This equation will represent the curve $omf$, where $f$ acts as a variable parameter. This equation describes countless curves $omf$, which extend infinitely along the axis $AO$, and eventually stabilize at a distance $f$ from it, except for the case when $f = 0$, which corresponds to the circle $AMF$.

As $X$ approaches infinity, $Y$ converges to $f$. However, not all of these curves represent actual streams, because each subsequent curve is not derived in the same manner from its predecessor, unlike the first curve, which is constructed from the circle itself.

In this passage, the author continues exploring the mathematical description of fluid flow around a quarter-circle. By defining a set of coordinates and equations, the author derives curves that represent possible paths for water streams near the body. The analysis highlights the complexity of fluid dynamics, showing how the stream paths vary based on their distance from the body and depend on specific parameters.

*XXV. Si curva AMF fuerit alia curva quae cunque, aequatione inter CP = x et PM = y contenta, et pro rivulo proximo omf ponatur Cp = X et PM = Y, erit Y = y + f et* $X = x - \frac{fdy}{dx}$, *siquidem intervallum f fuerit minimum. At quoniam figura sequentium rivulorum a praecedentibus simili modo definitur, si intervallum Ff = f statuatur finitum, curvae omf figura expressione magis complicata definietur. Ac pro applicata quidem erit Y = y + f, verum abscissa X talis erit functio ipsarum x et f, ut sit* $\frac{dy}{dx} + \left(\frac{dX}{dx}\right)\left(\frac{dX}{df}\right) = 0$, *unde natura functionis X determinatur. Si enim ponatur per seriem*

*$X = x - fP + ffQ - f^3R + f^4S - etc.$ existentibus P, Q, R, S etc. functionibus ipsius x, cuius quoque data est functio y, erit*

$$dy = (dx - fdP + f^2dQ - f^3dR + f^4dS\,etc.)(P - 2fQ + 3ffR - 4f^3S + etc.)$$

*unde sit:*

$$P = \frac{dy}{dx};\ Q = \frac{-PdP}{2dx};\ R = \frac{-PdQ - 2QdP}{3dx};\ S = \frac{-PdR - 2QdQ - 3RdP}{4dx}\ etc.$$

*sicque data curva AM omnes rivulorum curvae omf assignabuntur, ac per seriem quidem infinitam.*

XXV. If the curve $AMF$ is any arbitrary curve, defined by an equation relating $CP = x$ and $PM = y$, and for the closest stream $omf$, we let $Cp = X$ and $PM = Y$, then $Y = y + f$ and $X = x - \frac{f\,dy}{dx}$, assuming the interval $f$ is minimal.

Since the shape of subsequent streams is defined similarly to the preceding ones, if the interval $Ff = f$ is considered finite, the shape of the curve $omf$ will be determined by a more complicated expression. Specifically, the ordinate will be $Y = y + f$, but the abscissa $X$ will be a function of both $x$ and $f$, such that:

$$\frac{dy}{dx} + \left(\frac{dX}{dx}\right)\left(\frac{dX}{df}\right) = 0$$

This equation determines the nature of the function $X$. If we expand $X$ as a series:

$$X = x - fP + f^2Q - f^3R + f^4S - \text{etc.}$$

where $P, Q, R, S$, etc., are functions of $x$, and $y$ is also a given function of $x$, then:

$$dy = \left(dx - f\,dP + f^2\,dQ - f^3\,dR + f^4\,dS\,\text{etc.}\right)\left(P - 2fQ + 3f^2R - 4f^3S + \text{etc.}\right)$$

From this, we get:

$$P = \frac{dy}{dx};\quad Q = \frac{-P\,dP}{2\,dx};\quad R = \frac{-P\,dQ - 2Q\,dP}{3\,dx};\quad S = \frac{-P\,dR - 2Q\,dQ - 3R\,dP}{4\,dx};$$

Thus, given the curve $AM$, all the stream curves $omf$ can be assigned using an infinite series.

This section presents a generalization for calculating the paths of fluid streams around any arbitrary curved body. It introduces the use of infinite series to define these paths, allowing the computation of streamlines for complex shapes.

*XXVI. Quoniam hae formulae tantum vero proxime tractum singulorum rivulorum declarant, superfluum foret, in iis evoluendis operam consumere. Verae tamen formulae ab his non admodum erunt diversae, ac fortasse earum resolutio multo facilior evadet. Praeterea vero notari convenit, formulas veras non omnino determinatas esse posse, nisi forte extensio fluvii in latitudinem sit infinita; nam utcunque fluvius circa corpus cursum inflectat, ad ripam tamen eius directionem sequetur. Unde aequatio inter X et Y ita debet esse comparata, ut posito f = 0, praebeat ipsam corporis figuram AM; sin autem ipsi f certus quidam valor tribuatur,*

*ut tum figuram ipse exhibeat. Ita si ripa rectae AO ad distantiam = b fuerit paralela, ac ponatur CF = a, aequatio inter X et Y has 34roprietate habere debet, ut posito f = 0, inde ipsa curva AM resultet, sev fiat X = x et Y = y, sin autem ponatur f = b – a, quo casu punctum f in ripam cadet, ut tum fiat Y = b quicunque valor pro X sit proditurus.*

**XXVI.** Since these formulas only approximate the true paths of individual streamlines, it would be unnecessary to spend excessive effort in fully expanding them. However, the true formulas will not differ significantly from these, and perhaps their solution may even be simpler.

It should also be noted that the exact formulas cannot be fully determined unless the river extends infinitely in width. No matter how the river bends around the body, it will eventually follow the direction of the riverbank. Therefore, the equation between $X$ and $Y$ must be constructed in such a way that, when $f = 0$ is set, it gives the shape of the body $AM$; but if a certain value is assigned to $f$, it should represent the shape accordingly.

For example, if the riverbank is parallel to the straight line $AO$ at a distance $b$, and $CF = a$, the equation between $X$ and $Y$ should have the following properties: when $f = 0$, the curve $AM$ should result, i.e., $X = x$ and $Y = y$; but if $f = b - a$ is set, which makes the point $f$ fall on the riverbank, then $Y = b$ should result, regardless of the value of $X$.

In this section, the focus is on the limitations of the earlier formulas and the conditions under which they would accurately describe fluid flow around a body. The key point is the relationship between the streamlines and the boundary conditions, especially when the riverbank is considered.

*XXVII. Hinc autem satis probabiliter resistentiam definire poterimus, qua corpus AMF in fluido canali OCIH datae amplitudinis CH = b motum patitur, ad quem casum regula vulgaris non est accommodata. Sit igitur celeritas, qua corpus secundum direcionem AO promovetur, = c, et rivuli axi proximi amplitudo Oo = e; amplitudo autem corporis maxima CF = a; ut spatium in canili residuum sit FH = b - a, per quod cum fluidum omne defluere debeat, assumo enim, id neque supra corpus neque infra defluere posse, amplitudo rivuli in Ff erit* $\frac{b-a}{b}e = f$*, ubi celeritas debita sit altitudini k ut sit kff = cee, sev* $k = \frac{cbb}{(b-a)^2}$. *Ponatur nunc pro corporis figura CP = x; PM = y; et pro rivulo Cp = X et pm = Y, neque hic erit Y – y = f, neque Y - y = e, sed medium quendam tenebit valorem, ut sit* $Y - y = \frac{b-y}{b}e$. *At est* $Y - y : Mm = dx : \sqrt{(dx^2 + dy^2)}$*, unde* $Mm = \frac{b-y}{b} \cdot \frac{\sqrt{(dx^2+dy^2)}}{dx}$. *Si ergo celeritas aquae ad M defluentis debita sit altitudini v, erit* $\frac{(b-y)^2 ee(dx^2+dy^2)}{bbdx^2} v = cee$*, sev* $v = \frac{cbbdx^2}{(b-y)^2(dx^2+dy^2)}$.

**XXVII.** Hence, we can reasonably estimate the resistance that the body $AMF$ experiences when moving through a fluid in a channel $OCIH$ with a given width $CH = b$, a scenario for which the usual rule is not applicable. Let the velocity of the body in the direction $AO$ be $c$, and the amplitude of the nearest streamline to the axis be $Oo = e$. The maximum amplitude of the body is $CF = a$, leaving a remaining space in the channel of $FH = b - a$, through which all the fluid must flow. Assuming that the fluid cannot flow above or below the body, the amplitude of the streamline at $Ff$ will be $f = \frac{b-a}{b}e$, where the velocity corresponding to the height $k$ is such that $kff = cee$, or $k = \frac{cbb}{(b-a)^2}$.

Now, considering the shape of the body as described by the coordinates $CP = x$ and $PM = y$, and for the streamline $Cp = X$ and $pm = Y$, the difference $Y - y$ will neither be $f$ nor $e$, but will take an intermediate value such that $Y - y = \frac{b-y}{b}e$. The relationship is $Y - y : Mm = dx : \sqrt{dx^2 + dy^2}$, leading to $Mm = \frac{b-y}{b} \cdot \frac{\sqrt{dx^2+dy^2}}{dx}$.

Thus, if the velocity of the fluid at point $M$ corresponds to the height $v$, the equation becomes:

$$\frac{(b-y)^2 ee(dx^2 + dy^2)}{bb \cdot dx^2} v = cee,$$

or simplified,

$$v = \frac{cbb \cdot dx^2}{(b-y)^2 \cdot (dx^2 + dy^2)}.$$

This equation expresses how the fluid velocity at point $M$ is influenced by the geometry of the body and the constraints of the channel.

This section introduces a more complex scenario where the body is placed in a confined fluid channel, and the fluid must flow around the body within the given space. The resistance and velocity are determined based on the body's geometry and the channel's dimensions.

*XXVIII. Iam vero fluidi pressio in M est per resistentiae theoriam veram aequalis altitudini C— v. Sed quia in F pressio debet esse nulla, evidens est, fore* $C = k = \frac{cbb}{(b-a)^2}$, *unde pressio in M erit* $= \frac{cbb}{(b-a)^2} - \frac{cbbdx^2}{(b-y)^2(dx^2+dy^2)}$, *quae cum sit normalis ad corporis superficiem, inde nascetur resistentia ex curvae element* $\sqrt{(dx^2+dy^2)}$ *oriunda* $= -\frac{cbbdy}{(b-a)^2} + \frac{cbbdx^2dy}{(b-y)^2(dx^2+dy^2)}$, *cuius integrale dabit totam resistentiam. Si amplitudo fluid b esset infinita, foret resistentia* $= -cy + c\int \frac{dx^2dy}{dx^2+dy^2}$. *Si ergo AMF fuerit linea recta AF, sitque CA = b, existente CF = a; erit a − y:x = a:b, sev a − y =* $\frac{ax}{b}$, *et dy =-* $\frac{adx}{b}$, *hincque* $dx^2 + dy^2 = \frac{dx^2(aa+bb)}{bb}$. *Unde hoc casu resistentia erit* $= C + \frac{acbbx}{b(b-a)^2} + \frac{bbcbb}{(aa+bb)(b-y)}$ *quae per totam rectam AF extenta fiet:* $\frac{acbb}{(b-a)^2} + \frac{bbcb}{aa+bb} - \frac{bbcbb}{(aa+bb)(b-a)} = \frac{acbb}{(b-a)^2} - \frac{abbcb}{(aa+bb)(b-a)}$, *quae expressio abit in hanc:* $\frac{aacb(ab+bb)}{(aa+bb)(b-a)^2}$. *At*

*si fluidum esset infinitum, resistentia foret* $= \frac{a^3 c}{aa+bb}$*; quae si pouatur = R, illa resistentia erit* $= \frac{b(ab+bb)}{a(b-a)^2} R$*, ideoque maior, quam R.*

**XXVIII.** Now, according to the true theory of resistance, the pressure of the fluid at point M is equal to the height $C - v$. However, since the pressure at point F must be zero, it is evident that $C = k = \frac{cbb}{(b-a)^2}$. Thus, the pressure at M will be $\frac{cbb}{(b-a)^2} - \frac{cbb \cdot dx^2}{(b-y)^2(dx^2+dy^2)}$. Since this pressure is normal to the surface of the body, the resistance arising from the curve element $\sqrt{dx^2 + dy^2}$ will be $-\frac{cbb \cdot dy}{(b-a)^2} + \frac{cbb \cdot dx^2 \cdot dy}{(b-y)^2(dx^2+dy^2)}$, and integrating this will give the total resistance. If the fluid's width $b$ were infinite, the resistance would be $-cy + c\int \frac{dx^2 \cdot dy}{dx^2+dy^2}$.

If AMF were a straight line AF, with CA = b and CF = a, then $a - y : x = a : b$, or $a - y = \frac{ax}{b}$, and $dy = -\frac{adx}{b}$, hence $dx^2 + dy^2 = \frac{dx^2(aa+bb)}{bb}$. Therefore, in this case, the resistance would be:

$$C + \frac{acbbx}{b(b-a)^2} + \frac{bbcbb}{(aa+bb)(b-y)},$$

and extending this expression along the entire line $AF$, the total resistance will be:

$$\frac{acbb}{(b-a)^2} + \frac{bbcb}{aa+bb} - \frac{bbcbb}{(aa+bb)(b-a)} = \frac{aacb(ab+bb)}{(aa+bb)(b-a)^2}.$$

If the fluid were infinite, the resistance would be:

$$\frac{a^3c}{aa+bb},$$

and if we denote this by $R$, the resistance in a finite fluid will be:

$$\frac{b(ab+bb)}{a(b-a)^2}R,$$

which is greater than $R$.

In this paragraph, the author presents a mathematical formulation to calculate the resistance experienced by a body moving through a fluid within a confined channel. The approach relies on the true theory of resistance, which ties the pressure at a point on the body's surface to the velocity of the fluid at that point. The pressure distribution is used to derive an expression for the resistance, considering both finite and infinite fluid scenarios.

*XXIX. Experientia quoque hoc ipsum egregie confirmat, qua constat corpus in canali angustiori promotum, maiorem pati resistentiam, quam in canali ampliori, atque adeo si amplitudo corporis CF amplitudini canalis fuerit aequalis, ita ut corpus canalem perfecte expleat, tum resistentiam fieri infinitam. Quia enim fluidum non nisi per spatium FH defluere posse assumitur, hoc spatio evanescente corpus moveri non posset, quin fluidum in minus volumen compingeretur; at fluidum nullius compressionis capax assumitur. Quod si planum ad directionem motus fuerit normale, uti si ipfa linea CF = a celeritate* $\sqrt{c}$ *in directione CO promoveatur,*

*resistentia in fluido infinito foret = ac = R, in canali autem amplitudinis CH = b, eadem linea resistentiam sustinebit* $= \frac{bb}{(b-a)^2}R$, *quae ergo erit ad illam ut* $CH^2$ *ad* $FH^2$. *Nisi ergo amplitado CH prae amplitudine corporis CF fuerit praegrandis, augmentum resistentiae erit notabile. Sic si CH = 2CF erit resistentia. = 4R, si CH = 3CF , erit ea* $= \frac{9}{4}R$; *ac si fuerit CH = 10CF, erit resistentia* $\frac{100}{81}R = 1\frac{19}{81}R$.

**XXIX.** Experience also excellently confirms this, showing that a body moving in a narrower channel encounters greater resistance than in a wider channel. If the width of the body, CF, equals the width of the channel, so that the body completely fills the channel, the resistance becomes infinite. For, if it is assumed that the fluid can only flow through the space FH, then as this space vanishes, the body cannot move without compressing the fluid into a smaller volume. But it is assumed that the fluid is incompressible. If a plane is normal to the direction of motion, such that the line CF = a moves with velocity $\sqrt{c}$ in the direction CO, the resistance in an infinite fluid would be $ac = R$. In a channel of width CH = b, the same line will experience resistance $\frac{bb}{(b-a)^2}R$, which is to the former as $CH^2$ is to $FH^2$. Therefore, unless the channel's width CH is much larger than the width of the body CF, the increase in resistance will be significant. For instance, if CH = 2CF, the resistance will be $4R$; if CH = 3CF, it will be $\frac{9}{4}R$; and if CH = 10CF, the resistance will be $\frac{100}{81}R = 1\frac{19}{81}R$.

This section emphasizes the practical implications of the theory, showing that the geometry of the channel and the body's dimensions play a crucial role in determining the resistance encountered during motion.

*XXX. Quanquam autem hinc rivulorum, per quos aqua circa quodque corpus defluit, designatio non adeo difficilis videtur, tamen eorum natura cum principio continuitatis vix conciliari potest. Cum enim rivulorum partem corporis anticam cingentium amplitudo sit cosinui anguli, quem tangens corporis cum directione motus constituit, reciproce vero saltem proxime proportionalis, iuxta partem posticam vero eorum amplitudo sit quasi constans, nulla huius anguli, quem* $\phi$ *vocavimus, functio excogitari posse videtur, quae pro parte antica, ubi hic angulus est positivus, eius cosinum proxime exhibeat, pro parte autem postica; ubi iste angulus sit negativus, quasi non amplius ab hoc angulo pendeat, sed cons tans evadat. Interim hoc certum est, amplitudinem rivuli exacte per* $\frac{1}{\cos\phi}$ *non exprimi, quia tum similis mutatio circa partem posticam locum habere deberet, quod veritati repugnat. Causam quidem ampliationis rivulorum in parte antica agnoscimus, simulque in parte postica absentiam huius causae concedere debemus, sed quomodo haec cum principio continuitatis, cui calculus est superstruendus, cohaereant, nullo modo patet, ex quo summa difficultas, qua Theoria motus fluidorum etiam nunc premitur, multo magis perspicitur, quo propius ad eam pertingere videmur.*

**XXX.** Although the designation of the streamlets through which water flows around any body does not seem too difficult, their nature is hardly reconcilable with the principle of continuity. For the amplitude of the streamlets surrounding the front part of the body is inversely proportional—at least approximately—to the cosine of the angle formed by the tangent to the body and the direction of motion. However, near the rear part, the amplitude of the streamlets remains almost constant, and no function of this angle (which we have called $\phi$) seems conceivable that would provide the cosine for the front part, where the angle is positive, while for the rear part, where the angle is negative, it no longer depends on this angle and becomes constant. Meanwhile, it is certain that the amplitude of the streamlet is not exactly expressed by $\frac{1}{\cos\phi}$, because such a change should also occur around the rear part, which contradicts reality. We recognize the cause of the expansion of the streamlets in the front part and admit the absence of this cause in the rear part. However, how this aligns with the principle of continuity, which the calculus is built upon, is not at all clear. From this, the great difficulty with which the theory of fluid motion is still burdened becomes even more apparent, the closer we seem to approach it.

This paragraph highlights a key challenge in fluid dynamics: the behavior of streamlets around a moving body doesn't align well with the principle of continuity. The front part of the body shows a change in streamlet amplitude related to the cosine of the angle with the motion direction, but the rear part shows a constant amplitude, independent of this angle. This discrepancy complicates the theoretical understanding, showing that even though the theory has advanced, significant issues remain unresolved.

*XXXI. Quae hactenus tradita sunt, tantum ad resistentiam plani proprie sunt referenda, nihilo vero minus resistentia navis aliusue corporis in aqua moti inde colligi potest, dum eius partem submersam per sectiones inter se parallelas in strata minutissima sectam concipimus. Ita si AMENB fuerit sectio navis quaecunque horizontalis, eius resistentiam inde quoque aestimare licet, siquidem aqua resistens in hoc plano permaneat, neque sursum vel deorsum iuxta navem defluat. Quod igitur ad figuram puppis ENB attinet, in genere intelligimus, aquam iuxta eam defluere non posse, nisi lineae ENB curvatura sit ubique valde exigua. Cum enim in Ef nulla detur pressio, nulla inde vis adest, quae motum aquae ab E secundum directionem axi AB parallelam progressurae inflecat, atque hanc inflexionem a sola gravitate aquae produci debere, quod quidem in sectionibus profundioribus citius evenit, quam magis elevatis. Tum vero, quo velocius navis promovetur, eo difficilius aquae decursus incurvatur, et nisi inflexio ENB sit satis parva, aqua navem deferet, et ob deficientem ibi pressionem aquae, resistentia prorae etiam a pondere aquae proram urgente augebitur, quod ingens vitium navium reputatur.*

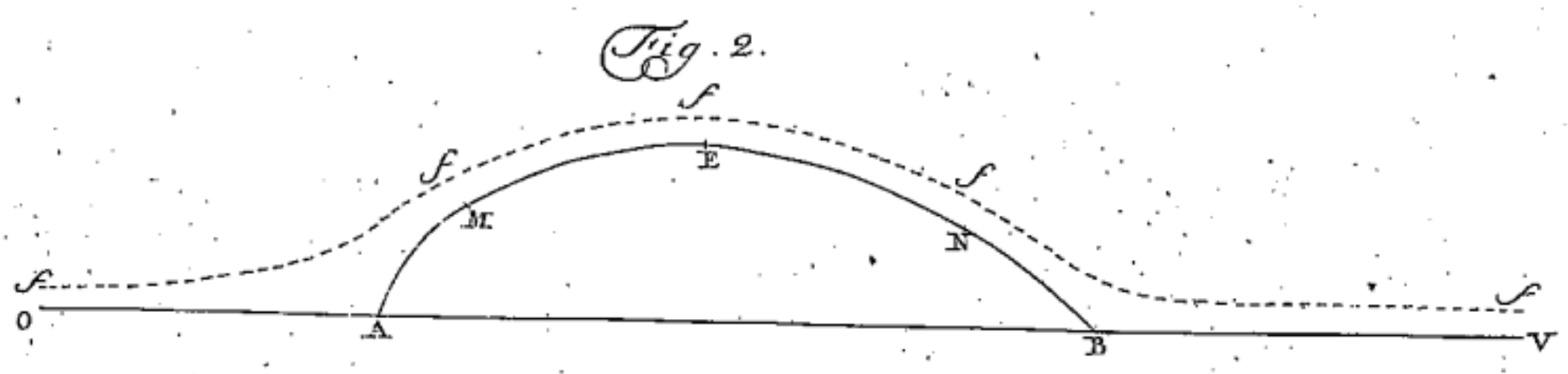

XXXI. What has been discussed so far is primarily related to the resistance of a plane. However, the resistance of a ship or any other body moving in water can also be deduced from this, by imagining its submerged part as being divided into very thin parallel layers. Thus, if AMENB represents a horizontal cross-section of any ship, its resistance can also be estimated in this manner, provided that the water resisting it remains on this plane and does not flow upwards or downwards along the ship's sides. Regarding the shape of the stern ENB, we generally understand that water cannot flow around it unless the curvature of the line ENB is very small everywhere. Since there is no pressure at Ef, there is no force to deflect the water from its motion along the axis AB, and any such deflection must be caused solely by the water's gravity. This occurs more quickly in deeper sections than in higher ones. Additionally, the faster the ship moves, the harder it is for the water to curve around, and if the curvature of ENB is not small enough, the water will be carried along with the ship. Due to the lack of pressure in that area, the resistance at the bow will be increased by the water's weight pressing against it, which is considered a significant flaw in ship design.

This passage applies the principles of fluid resistance to the design of ships. By imagining the submerged part of a ship as being divided into thin layers, the author explains that resistance can be estimated similarly to how it is for a plane. The key point here is the curvature of the ship's stern. If the curvature is too steep, the water cannot flow smoothly around it, leading to increased resistance. This is especially problematic as the ship's speed increases, and it results in additional pressure at the bow due to the weight of the water, which is considered a major flaw in ship design. The passage underscores the importance of minimizing curvature in ship design to reduce resistance and improve efficiency.

*XXXII. Etiamsi autem aqua iuxta puppem ENB bene defluat, neque istud incommodum sit pertimescendum, tamen hoc ad facilem gubernaculi actionem, ad quam non minus navem instructam esse oportet, non sufficit. Cum enim aqua fere usque ad B defluxerit; quia ab altera parte simili modo fertur, perinde motum continuare debet, quasi secundum rectam BV obex ipsi obiiceretur, et quia prope B cursum inflectere cogitur, perinde uti in A est fatcum, eius motus eo magis retardabitur, quo maior fuerit angulus ABN, quod quidem in maioribus navibus viu venire potest, etiam si linca ENB sit arcus circuli admodum magni; in aqua autem circa B fere stagnante gubernaculum vix ullam vim exerere valebit. Quocirca necesse est, ut figura ENB non solum lente incurvetur, sed etiam in B cum axe AB angulum satis acutum constituat. Interim tamen, ob istam aquae retardationem circa B, navis inde maiorem pressionem sustinebit, qua resistentia prorae imminuetur, unde, nisi gubernaculi ratio haberi deberet, angulus fere rectus ad B cursum navis potius acceleraret, quam retardaret.*

**XXXII.** Even if the water flows smoothly along the stern ENB and the previously mentioned issue is not a concern, this alone is not sufficient for the proper functioning of the rudder, which is equally important for a well-equipped ship. When the water has nearly reached point B, since it flows similarly from the other side, it must continue its motion as if an obstacle were placed along the line BV. As the water is forced to change direction near point B, just as it was at point A, its movement will be increasingly hindered the larger the angle ABN is. This can happen in larger ships even if the line ENB forms a very large circular arc. Additionally, with the water near point B nearly stagnant, the rudder will have very little effect. Therefore, it is necessary for the shape of ENB not only to curve gradually but also to form a sufficiently acute angle with the axis AB at point B. However, due to the slowing of the water near point B, the ship will experience greater pressure, which reduces the resistance at the bow. Therefore, unless the rudder is taken into account, a nearly right angle at point B would accelerate rather than hinder the ship's progress.

This passage addresses the importance of stern design and its impact on both rudder effectiveness and overall ship resistance. The author emphasizes that even with an optimized stern curvature, the angle at point B (where the water flows off the stern) is crucial for maintaining control and minimizing resistance. If the angle is too wide, it can cause water stagnation, reducing the rudder's effectiveness and increasing resistance. However, a sharper angle at point B helps maintain water flow and reduces pressure on the bow, potentially improving the ship's speed. The passage highlights the delicate balance between reducing resistance and ensuring effective steering in ship design.

*XXXIII. Hae autem considerationes ad commodiorem evolutionem formularum, quibus universa Theoria motus fluidorum continetur, viam aperire videntur. Cum enim istae formulae in genere pro quocunque loco tam motum fluidi, quam pressionem, exhibeant, quae summa generalitas in causa erat, quod hae formulae minus tractabiles evaserint, ea, quae hactenus sunt allata, non exiguam spem facilioris calculi faciunt, si non folum rivulos, per quos singulae aquae particulae deferuntur, contemplemur, sed etjam harum curvarum traiectorias orthogonales in calculum introducamus: quoaiam enim hae traiectoriae cuiusque rivuli in quojue loco amplitudiaem commodissime ostendunt , inde celeritas aquae, quae in quolibet rivulo amplitudini reciproce est proportionalis, aptissime definitur, unde deinceps pressio per formulam concinniorem exprimi posse videtur. Assumo autem, tam omnem aquam, quam eius motum, in eodem plano esse constitutum, eumque iam ita ad statum permanentem esse perductum, ut rivulorum tractus sint constantes, neque ulli amplius mutationi obnoxiae.*

**XXXIII.** These considerations seem to open the way for a more convenient development of the formulas that encompass the entire theory of fluid motion. Since these formulas generally describe both the motion of the fluid and the pressure at any given location, this generality has been the reason why these formulas have become less manageable. However, what has been presented thus far offers a significant hope for easier calculations, especially if we not only consider the streamlines along which each water particle is transported but also introduce the orthogonal trajectories of these curves into the calculation. These trajectories conveniently show the amplitude of each streamline at any given location. Thus, the velocity of the water, which is inversely proportional to the amplitude of any streamline, is more accurately defined, and consequently, the pressure can be expressed by a more concise formula. I assume, however, that all the water and its motion are confined to the same plane and that the system has already reached a steady state, where the paths of the streamlines are constant and subject to no further changes.

In this passage, the author proposes a method for simplifying the complex equations governing fluid motion. By focusing on streamlines and their orthogonal trajectories, which provide insights into the amplitude and velocity of the fluid, the author suggests that the pressure and other variables in fluid dynamics could be more easily calculated. This approach aims to address the difficulties in managing the general formulas and offers a more structured and potentially more practical way to approach problems in fluid dynamics, particularly in steady-state conditions. The emphasis on orthogonal trajectories as a key tool in simplifying the calculations marks an important shift in the methodology for solving fluid dynamics problems.

*XXXIV. Quo autem haec facilius ad Theoriam resistentiae accommodari queant, omnes determinationes ad figuram corporis AME aquae immissi referri conveniet. Hanc ergo figuram pro fixa habebo, quia in resistentiae investigatione perinde est, sive corpus contra aquam stagnantem, sive aqua contra corpus quiescens pari celeritate feratur. luxta corpus ergo aqua, quicunque motus ei tribuatur, secundum eius figuram AME praeterfluet, et in maioribus distantiis motus aquae fiet per certas lineas curvas RYS, rys, quibus rivuli constituuntur. Talium rivulorum series intra AME et RYS infinita multitudo concipi debet, quae omnes inter se tantum ratione parametri differant. Sit MYy traiectoria orthogonalis quaecunque, quae ex M egressa omnes rivulos normaliter traiiciat, uti etiam in M ad ipsam curvam datam AME est normalis. Hocque modo puncta rivulorum Y et y inprimis cum puncto M connectuntur, ut magis ad boc punctum, quam ad aliud quodvis pertinere sint censenda.*

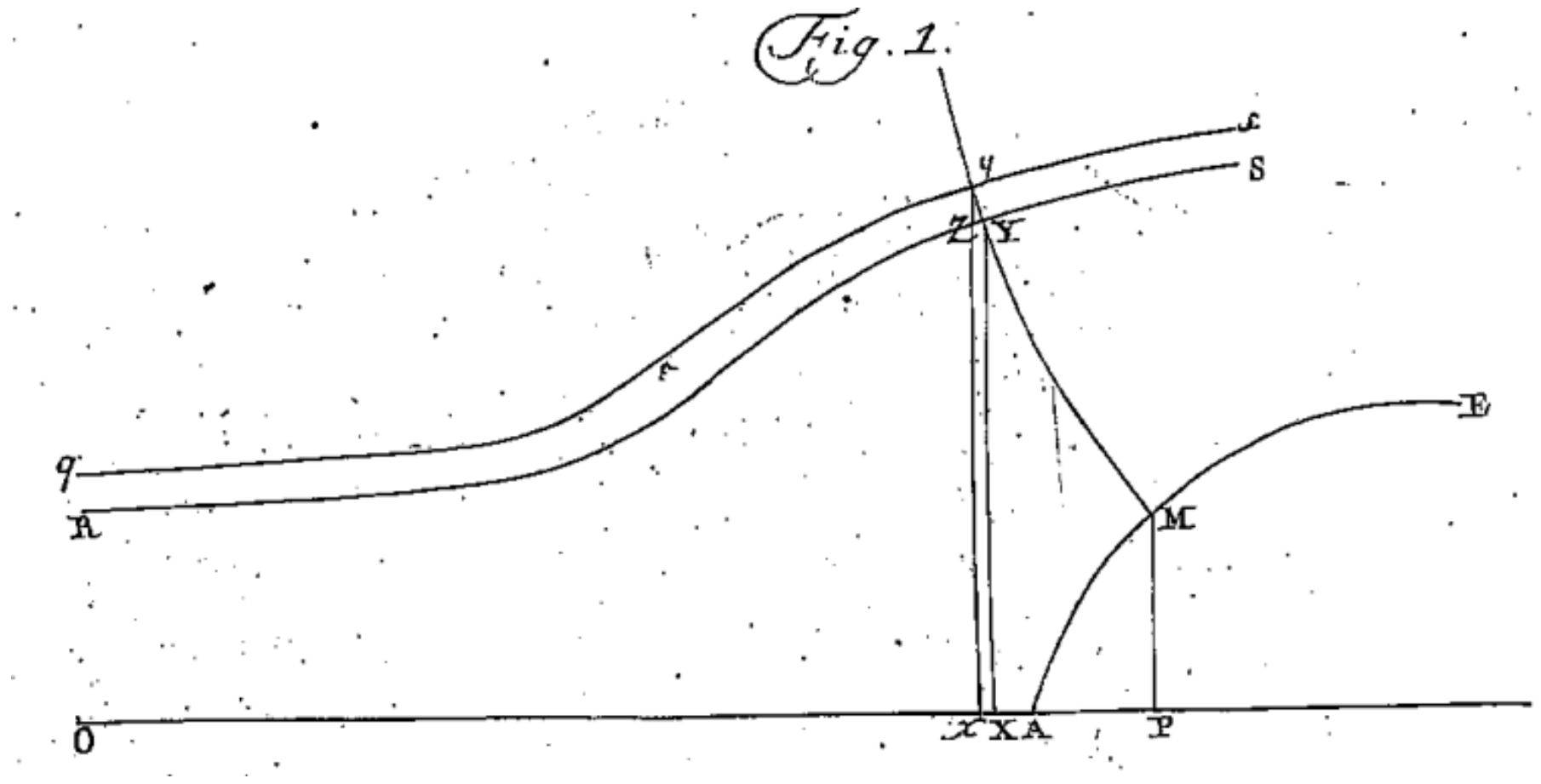

XXXIV. To make this more easily applicable to the theory of resistance, it is convenient to refer all determinations to the shape of the body AME immersed in water. Therefore, I will consider this shape as fixed because, in the investigation of resistance, it makes no difference whether the body moves against stationary water or the water moves against a stationary body at the same speed. Thus, the water flows past the body according to its shape, AME, regardless of the motion attributed to it. At greater distances, the motion of the water will follow certain curved lines RYS, rys, along which streamlines are formed. An infinite number of such streamlines should be conceived between AME and RYS, all differing only in their parameters. Let MYy be any orthogonal trajectory, which, starting from M, intersects all the streamlines normally, just as it is also normal to the given curve AME at M. In this way, the points of the streamlines Y and y are primarily connected with point M, so they are considered more related to this point than to any other.

In this section, the author discusses how to apply the theory of resistance by focusing on the shape of a body immersed in water. By assuming the shape is fixed, the analysis can be simplified whether the body moves through stationary water or the water flows past a stationary body. The author introduces the concept of streamlines, which are curved paths that the fluid follows around the body. He emphasizes the importance of orthogonal trajectories, which intersect the streamlines at right angles, to better understand the fluid's behavior around the body. This approach aims to simplify the complex calculations related to fluid resistance.

*XXXV. Ponamus ergo pro isto puncto M abscissam AP = s, et applicata PM exprimetur per certam quandam functionem ipsius s: pro rivulo autem RYS parameter sit = b, qui pro sequente rys abeat in b + db, pro ipsa autem curva AME evanescat. Iam situs puncti Y pendebit partim a puncto M, partim a parametro, unde eius coordinatae, quae sint AX = x, XY = y, erunt functiones istarum duarum quantitatum set b; ponamus ergo:*

$$dx = Pds + Qdb \text{ and } dy = Rds + Sdb,$$

*quae relatio inter x et y ita debet esse comparata, ut, posito b = 0, ipsam curvam AME praebeat: at si ipsi b certus quidem et constans valor tribuatur, aequatio sit proditura pro curuva RYS; pro qua ergo erit dx = Pds et dy = Rds. Sin autem punctum M fixum sumatur, variabilitas solius parametri b dabit traiectoriam orthogoalem*

*MYy, pro qua ergo ducta applicata proxima xy, et Yz, axi AX parallela, erit Xx = Qdb et yz = Sdb; quia pro punctis in eadem traiectoria sitis quantitas s non variatur.*

**XXXV.** Let us assume that for this point M, the abscissa AP is denoted by $s$, and the ordinate PM is expressed as a certain function of $s$. For the streamline RYS, let the parameter be $b$, which changes to $b + db$ for the next streamline $rys$, and becomes zero for the curve AME. The position of point Y will depend partly on point M and partly on the parameter $b$, so its coordinates, $AX = x$ and $XY = y$, will be functions of these two quantities $s$ and $b$. Therefore, we assume: $dx = Pds + Qdb \quad \text{and} \quad dy = Rds + Sdb$, where the relationship between $x$ and $y$ must be such that when $b = 0$, it yields the curve AME. If a fixed value is assigned to $b$, the equation will produce the curve RYS, for which $dx = Pds$ and $dy = Rds$. However, if point M is fixed, the variation of the parameter $b$ alone will give the orthogonal trajectory MYy, for which the next parallel ordinates $xy$ and $Yz$, parallel to the axis $AX$, will be given by $Xx = Qdb$ and $yz = Sdb$, since the quantity $s$ does not vary for points on the same trajectory.

This paragraph introduces a mathematical framework to describe the position of points along streamlines in a fluid flow. The author introduces coordinates $s$ and $b$, where $s$ represents the abscissa of a point on the curve AME and $b$ is a parameter that distinguishes different streamlines. The differential equations describe how the coordinates $x$ and $y$ change with respect to $s$ and $b$. The focus is on understanding how variations in the parameter $b$ affect the shape of streamlines and their orthogonal trajectories, which helps in the analysis of fluid resistance around a submerged body.

*XXXVI. Cum iam Yy sit id curvam RYS normalis, erit ex natura traiectoriarum orthogonalium zy:Yz = dx:-dy = P:-R unde sit S:Q = P:-R ideoque PQ + RS = 0. Ut huic conditioni satisfaciamuus, ponamus statim:*

*Q = RT et S = - PT ut sit*
*dx = Pds +RTdb et dy = Rds – PTdb.*

*Porro autem erit rivuli amplitudo Yy = db*$\sqrt{(QQ+SS)}$ = *Tdb*$\sqrt{(PP+RR)}$, *cui cum celeritas aquae in Y, quatenus aqua in eodem rivulo comparatur, sit reciproce proportionalis, posita celeritate in* $Y = \text{ȣ}$, *statuamus* $\text{ȣ} = \frac{B}{T\sqrt{PP+RR}}$, *ubi B denotat functionem ipsius parametri b tantum. Resolvatur haec celeritas secundum directiones coordinatarum x et y, sintque celeritates derivatae secundum AX = u et secundum XY = v; ac reperitur*

$$u = \frac{BP}{T(PP+RR)} \; et \; v = \frac{BR}{T(PP+RR)}$$

*unde ob uu + vv = ȣ ȣ erit ȣ ȣ =* $\frac{BB}{TT(PP+RR)}$.

**XXXVI.** Since $Yy$ is normal to the curve $RYS$, from the nature of orthogonal trajectories, we have $zy : Yz = dx : -dy = P : -R$, leading to the ratio $S : Q = P : -R$ and therefore $PQ + RS = 0$. To satisfy this condition, we set: $Q = RT$ and $S = -PT$, so that: $dx = Pds + RTdb$ and $dy = Rds - PTdb$. Furthermore, the amplitude of the streamline $Yy$ is given by: $Yy = db\sqrt{(Q^2+S^2)} = Tdb\sqrt{(P^2+R^2)}$, where the velocity of the water at $Y$, relative to the streamline, is inversely proportional to this amplitude. Assuming the velocity at $Y$ is $\mathfrak{v}$, we set: $\mathfrak{v} = \frac{BT}{\sqrt{P^2+R^2}}$, where $B$ is a function dependent only on the parameter $b$. Resolving this velocity into components along the $x$ and $y$ coordinates, the derived velocities are: $u = \frac{BP}{T(P^2+R^2)}$ and $v = \frac{BR}{T(P^2+R^2)}$. Since $u^2 + v^2 = \mathfrak{v}^2$, it follows that: $\mathfrak{v}^2 = \frac{B^2}{T^2(P^2+R^2)}$.

In this passage, Euler presents the relationship between the trajectory of the fluid's motion and its orthogonal curves. He defines the velocity components in terms of geometric properties and establishes that the velocity at any point on the streamline is inversely proportional to the amplitude of that streamline. This mathematical framework simplifies the calculation of fluid motion and resistance.

*XXXVII. Quia igitur est* $\frac{B}{T(PP+RR)} = \frac{T\mathfrak{v}\mathfrak{v}}{B}$, *habebimus:*

$$u = \frac{PT\mathfrak{v}\mathfrak{v}}{B} \; et \; v = \frac{RT\mathfrak{v}\mathfrak{v}}{B}.$$

*Conveniet autem potius ipsas has celeritates* $u$ *et* $v$ *in calculum introduci, quam quantitates P et R ibi relinqui, unde colligetur:*

$$P = \frac{Bu}{T\mathfrak{v}\mathfrak{v}};\; R = \frac{Bv}{T\mathfrak{v}\mathfrak{v}};\; Q = \frac{Bv}{\mathfrak{v}\mathfrak{v}};\; S = \frac{-Bu}{\mathfrak{v}\mathfrak{v}}$$

$$et\; dx = \frac{B}{T\mathfrak{v}\mathfrak{v}}(uds + Tvdb) \; et \; dy = \frac{B}{T\mathfrak{v}\mathfrak{v}}(vds - Tudb)$$

*quas ergo formulas integrabiles esse oportet. Quare quia* $\mathfrak{v}\mathfrak{v} = uu + vv$ *et B functio ipsius b tantum, facile colligitur, cuiusmodi functiones esse debeant u, v et T, ut his duobus requisitis satisfiat. Siquidem, quod regula vulgaris exigebat, celeritas in quovis rivulo proportionalis esset cosinui anguli, quem curva cum axe facit, sev* $\mathfrak{v} = \frac{CP}{\sqrt{PP+RR}}$, *haberemus Cu=*$\mathfrak{v}\mathfrak{v}$*, existente C functione ipsius b tantum, ideoque* $v = \sqrt{(Cu - uu)}$, *ita sev* $u = \frac{\mathfrak{v}\mathfrak{v}}{C}$ *et v=*$\frac{\mathfrak{v}}{C}\sqrt{(CC - \mathfrak{v}\mathfrak{v})}$, *ita ut integrabiles esse de berent hae formulae:*

$$dx = \frac{B}{CT}ds + \frac{B}{C\mathfrak{v}}db\sqrt{Cu - uu};\; dy = \frac{B}{CT}ds + \frac{Bds}{CT\mathfrak{v}}\sqrt{CC - \mathfrak{v}\mathfrak{v}} - \frac{B}{C}db.$$

**XXXXVII.** Since $\frac{B}{T(PP+RR)} = \frac{Tȣȣ}{B}$, we have: $u = \frac{PTȣȣ}{B}$ and $v = \frac{RTȣȣ}{B}$. It is preferable to introduce these velocities $u$ and $v$ into the calculation, rather than leaving the quantities $P$ and $R$ in place. From this, it follows that: $P = \frac{Bu}{Tȣȣ}$; $\quad R = \frac{Bv}{Tȣȣ}$; $\quad Q = \frac{Bv}{ȣȣ}$; $\quad S = \frac{-Bu}{ȣȣ}$ and: $dx = \frac{B}{Tȣȣ}(uds + Tvdb)$ and $dy = \frac{B}{Tȣȣ}(vds - Tudb)$.

These formulas must be integrable. Since $ȣȣ = uu + vv$ and $B$ is a function only of $b$, it is straightforward to determine the functions $u$, $v$, and $T$ that must satisfy these two requirements. According to the usual rule, the velocity in any stream should be proportional to the cosine of the angle the stream makes with the axis, i.e., $ȣ = \frac{CP}{\sqrt{PP+RR}}$. We then have: $Cu = ȣȣ$ where $C$ is a function only of $b$. Therefore: $v = \sqrt{Cu - uu}$ Thus: $u = \frac{ȣȣ}{C}$ and $v = \frac{ȣ}{C}\sqrt{CC - ȣȣ}$ The integrable formulas then become: $dx = \frac{B}{CT}ds + \frac{B}{Cȣ}db\sqrt{Cu - uu}$; $dy = \frac{B}{CT}ds + \frac{B}{CTȣ}ds\sqrt{CC - ȣȣ} - \frac{B}{C}db$.

> This section deals with the reformulation of certain mathematical expressions to simplify the calculation of stream velocities in a given system. By substituting $P$ and $R$ with $u$ and $v$, the formulas for $dx$ and $dy$ are derived, making them easier to integrate. The transformation to $u$ and $v$ helps in aligning with the requirement that velocities are proportional to the cosine of the angle with the axis. The final expressions for $dx$ and $dy$ are structured to be integrable, adhering to the conditions imposed by the problem. This approach illustrates the process of simplifying complex mathematical relationships to make them more tractable for integration.

*XXXVIII. Verum iam perpendamus, quid Theoria motus fluidorum requirat. Ostendi autem, si pressio aquae in Y exponatur per alitudinem p, et ex viribus acceleratricibus nascatur efficacia = V, tum sumtis x et y utcunque variabilibus; hanc aequationem locum habere debere:*

$$p = V - \int udx\frac{du}{dx} + vdx\frac{du}{dy} + udy\frac{dv}{dx} + vdy\frac{du}{dy}$$

*Totum ergo negotium huc redit, ut ista formula integrationem actu admittat; nisi enim hoc eveniat, talis motus, qualis per quantitates u et v singitur, omnino subsistere nequit. Si quaestio de pressione restringatur ad unicum rivulum, ostendi hoc integrale eo reduci, ut fiat p = V - $\frac{1}{2}$ȣȣ, ubi $\frac{1}{2}$ȣȣ referat altitudinem celeritati aquae debitam, uti iam supra inveni. Verum pro tota motus extensione necesse est, ut illud differentiale, cuius integrale occurrit, sit completum, uti quidem loquendi mos est.*

**XXXVIII.** But now, let us consider what the theory of fluid motion requires. It has been shown that if the pressure of water at $Y$ is represented by the height $p$, and if the accelerating forces give rise to an effect $V$, then, taking $x$ and $y$ as arbitrary variables, the following equation must hold:

$$p = V - \int \left( u\,dx\,\frac{du}{dx} + v\,dx\,\frac{du}{dy} + u\,dy\,\frac{dv}{dx} + v\,dy\,\frac{du}{dy} \right)$$

Thus, the entire task comes down to ensuring that this formula is integrable in practice; for if this does not happen, such motion, as described by the quantities $u$ and $v$, cannot exist at all. If the question of pressure is restricted to a single stream, I have shown that this integral reduces to:

$$p = V - \frac{1}{2}\text{ȢȢ}$$

where $\frac{1}{2}$ȢȢ represents the height due to the velocity of the water, as I have already found above. However, for the complete extension of motion, it is necessary that the differential, whose integral occurs, be complete, as is indeed the customary way of speaking.

This passage discusses the conditions required for a valid theory of fluid motion. The equation relates water pressure to various differential terms involving velocity components $u$ and $v$. The main concern is that for the described fluid motion to be physically possible, the equation must be integrable. The author points out that for a single stream, the pressure equation simplifies, but for the general motion, the differential terms must be complete to ensure consistency. This highlights the importance of mathematical consistency in the theoretical description of fluid dynamics.

*XXXIX. Quodsi formulas hactenus inventas huc transferre velimus, habemus quidem valores pro dx et dy; verum pro formulis $\frac{du}{dx}$ et $\frac{dv}{dy}$ notandum est in differentiatione ita solum x poni variabile, ut y maneat ivariatum; ergo ob dy = 0 erit Tudb = vds sev $db = \frac{vds}{Tu}$; unde sit $dx = \frac{Bds}{Tu}$. Quare si ponamus*

*du = Kds + Ldb et dv = Mds + Ndb*

*Erit in hac hypothesi*

$$\frac{du}{dx} = \left(Kds + \frac{Lvds}{Tu}\right) : \frac{Bds}{Tu} = \frac{KTu + Lv}{B}$$
$$\frac{dv}{dx} = \left(Mds + \frac{Nvds}{Tu}\right) : \frac{Bds}{Tu} = \frac{Mtu + Nv}{B}$$

*Similiter pro formulis $\frac{du}{dx}$ et $\frac{dv}{dy}$ assumitur x constans, unde erit $db = \frac{-uds}{Tv}$ et $dy = \frac{Bds}{Tv}$, sicque prodibit:*

$$\frac{du}{dy} = \left(Kds - \frac{Luds}{Tv}\right) : \frac{Bds}{Tv} = \frac{KTv - Lu}{B}$$
$$\frac{dv}{dy} = \left(Mds - \frac{Nuds}{Tv}\right) : \frac{Bds}{Tv} = \frac{MTv - Nv}{B}.$$

**XXXIX.** If we want to apply the formulas obtained so far to this context, we do indeed have the values for $dx$ and $dy$. However, for the formulas $\frac{du}{dx}$ and $\frac{dv}{dy}$, it should be noted that in differentiation, $x$ is treated as the only variable, while $y$ remains unchanged; therefore, since $dy = 0$, we will have $Tudb = vds$, or equivalently, $db = \frac{vds}{Tu}$. Thus, we get $dx = \frac{Bds}{Tu}$.

Therefore, if we assume:

$$du = Kds + Ldb \quad \text{and} \quad dv = Mds + Ndb,$$

we will have, under this assumption:

$$\frac{du}{dx} = \left(Kds + \frac{Lvds}{Tu}\right) : \frac{Bds}{Tu} = \frac{KTu + Lv}{B},$$

$$\frac{dv}{dx} = \left(Mds + \frac{Nvds}{Tu}\right) : \frac{Bds}{Tu} = \frac{MTu + Nv}{B}.$$

Similarly, for the formulas $\frac{du}{dy}$ and $\frac{dv}{dy}$, assuming $x$ is constant, we will have $db = \frac{-uds}{Tv}$ and $dy = \frac{Bds}{Tv}$. Thus, it will follow that:

$$\frac{du}{dy} = \left(Kds - \frac{Luds}{Tv}\right) : \frac{Bds}{Tv} = \frac{KTv - Lu}{B},$$

$$\frac{dv}{dy} = \left(Mds - \frac{Nuds}{Tv}\right) : \frac{Bds}{Tv} = \frac{MTv - Nv}{B}.$$

This passage presents the differentiation of the velocity components $u$ and $v$ with respect to $x$ and $y$, assuming certain conditions. The author introduces the variables $K$, $L$, $M$, and $N$ to represent the derivatives of $u$ and $v$. The formulas are adjusted based on the assumption that $y$ or $x$ is constant during differentiation, leading to expressions for the partial derivatives of the velocity components. This approach is essential for analyzing fluid motion in a more detailed mathematical framework.

*XL. Ex his ergo differentiale superius, cuius integrale in formulam pro p datam ingreditur, abibit in formam sequentem:*

$$+\frac{u}{T88}(uds + Tvdb)(KTu + Lv)$$

$$+\frac{v}{T88}(uds + Tvdb)(KTv - Lu)$$

$$+\frac{u}{T88}(vds - Tudb)(MTu + Nv)$$

$$+\frac{u}{T88}(vds - Tudb)(MTv - Nu)$$

*quae quatuor formulae statim has duas rediguntur:*

$$K(uds + Tvdb) + (vds - Tudb)$$

*Cum iam sit* $K = \frac{du}{dx}$ *et* M= $\frac{dv}{dx}$, *pressio quaesita sequenti definietur aequatione:*

$$p = V - \int (uds + Tvdb)\left(\frac{du}{ds}\right) + u(vds - Tudb)\left(\frac{dv}{ds}\right)$$

*sev ob* $udu + vdv = ȣdȣ$ *habebitur:*

$$p = V - \int \left( ds\left(\frac{ȣdȣ}{ds}\right) + Tdb\left(\frac{vdu - udv}{ds}\right)\right)$$

XL. Therefore, the differential mentioned above, whose integral enters into the formula for *p*, will transform into the following form:

$$+\frac{\mathrm{u}}{\mathrm{T}ȣȣ}(\mathrm{uds} + \mathrm{Tvdb})(\mathrm{KTu} + \mathrm{Lv})$$

$$+\frac{\mathrm{v}}{\mathrm{T}ȣȣ}(\mathrm{uds} + \mathrm{Tvdb})(\mathrm{KTv} - \mathrm{Lu})$$

$$+\frac{\mathrm{u}}{\mathrm{T}ȣȣ}(\mathrm{vds} - \mathrm{Tudb})(\mathrm{MTu} + \mathrm{Nv})$$

$$+\frac{\mathrm{u}}{\mathrm{T}ȣȣ}(\mathrm{vds} - \mathrm{Tudb})(\mathrm{MTv} - \mathrm{Nu})$$

These four formulas are immediately reduced to the following two:

$$\mathrm{K}(\mathrm{uds} + \mathrm{Tvdb}) + (\mathrm{vds} - \mathrm{Tudb})$$

Since $\mathrm{K} = \frac{\mathrm{du}}{\mathrm{dx}}$ et $\mathrm{M} = \frac{\mathrm{dv}}{\mathrm{dx}}$, the required pressure $p$ iis defined by the following equation:

$$= \mathrm{V} - \int (\mathrm{uds} + \mathrm{Tvdb})\left(\frac{\mathrm{du}}{\mathrm{ds}}\right) + \mathrm{u}(\mathrm{vds} - \mathrm{Tudb})\left(\frac{\mathrm{dv}}{\mathrm{ds}}\right)$$

Or, given that $\mathrm{udu} + \mathrm{vdv} = ȣdȣ$, we get:

$$\mathrm{p} = \mathrm{V} - \int \left( \mathrm{ds}\left(\frac{ȣdȣ}{\mathrm{ds}}\right) + \mathrm{Tdb}\left(\frac{\mathrm{vdu} - \mathrm{udv}}{\mathrm{ds}}\right)\right).$$

In this section, the author derives a more specific formula for pressure $p$ by combining velocity components with their derivatives and expressing them in a simplified form. The expression for pressure depends on integrating terms involving the differentials of velocities $u$ and $v$, and their products. The use of the relation $udu + vdv = ȣdȣ$ allows further simplification of the pressure formula. This result is crucial for understanding how pressure relates to fluid motion, particularly in systems where velocity components vary along different directions.

*XLI. Haec formula adhuc concinnior reddi potest, introducendo praeter ipsam celeritatem* ȣ, *eius quoque directionem. Sit ergo* $\phi$ *angulus quem directio motus in Y cum axe Ao facit, et quia sit* $u = ȣ \cos\phi$ *et* $v = ȣ \sin\phi$ *D, conficitur hinc* $u = vdu - udv = -ȣȣd\phi$, *sicque pressio p definietur per hanc aequationem:*

$$p = V - \int \left( ds\left(\frac{ȣdȣ}{ds}\right) - Tȣȣdb\left(\frac{d\phi}{ds}\right)\right)$$

*quae non amplius pendet a positione coordinatarum; utpote arbitraria; et hic quantitates,* $ȣȣ$ *et* $d\phi$ *considerandae sunt tanquam fünctiones ipsarum s et b. Hic vero evidens est, si b sumatur constans, integrationem nulam habere difficultatem, cum prodeat*

$$p = V - \frac{1}{2}ȣȣ + D$$

*denotante D funcionem parametri D; quare si et eius variabilitatis ratio habeatur, esse oportet*

$$db\left(\frac{ȣdȣ}{db}\right) = -Tȣȣdb\left(\frac{d\phi}{ds}\right)$$

*unde hoc obtinemus requisitum, ut esse debeat*

$$\frac{dD}{db} = \left(\frac{ȣdȣ}{db}\right) + Tȣȣdb\left(\frac{d\phi}{ds}\right) = functioni\ ipsius\ b\ tantum.$$

XLI. This formula can be made even more elegant by introducing, in addition to the velocity ȣ, its direction as well. Let ϕ be the angle that the direction of motion in Y makes with the axis Ao and let u=ȣcosϕ and v=ȣsinϕ. From this, $u = vdu - udv = -ȣȣd\phi$, thus the pressure p is defined by the following equation:

$$p = V - \int\left(ds\left(\frac{ȣdȣ}{ds}\right) - Tȣȣdb\left(\frac{d\phi}{ds}\right)\right)$$

This expression no longer depends on the arbitrary positioning of the coordinates. Here, the quantities ȣȣ and dϕ should be considered as functions of s and b. It is evident that if b is considered constant, the integration presents no difficulty, resulting in:

$$p = V - \frac{1}{2}ȣȣ + D$$

denoting D as a function of the parameter D. Therefore, if its variability is taken into account, it must be that:

$$\frac{dD}{db} = \left(\frac{ȣdȣ}{db}\right) + Tȣȣdb\left(\frac{d\phi}{ds}\right) = \text{a function of b only.}$$

In this section, the author refines the previously derived formula by incorporating the direction of velocity. By introducing the angle $\phi$, the formula becomes more general and independent of specific coordinate systems. The text highlights how, under certain conditions, the integration simplifies, leading to a straightforward expression for pressure $p$. The formula elegantly combines velocity magnitude and direction, showing the connection between the changes in the system and the pressure distribution.

*XLII. Verum insuper necesse est, ut formulae differentiales pro dx et dy inventae fiant completae sev integrabiles; valoribus autem pro u et v substitutis habemus:*

XLII. However, it is furthermore necessary that the differential formulas for dx and dy be made complete or integrable. After substituting the values for u and v, we have:

$$dx = \frac{B}{T\text{ȣ}}(ds\cos\phi + \mathrm{Tdb}\sin\phi)$$

$$dy = \frac{B}{T\text{ȣ}}(ds\sin\phi - \mathrm{Tdb}\cos\phi).$$

The formulas, in order to be integrable, must satisfy: if for the sake of brevity we let $\frac{1}{T\text{ȣ}} = \Theta$ ...

In paragraph XLII, Euler emphasizes the necessity of making the differential formulas for $dx$ and $dy$ integrable (i.e., ensuring that the equations can be solved through integration). After substituting the expressions for the velocity components $u$ and $v$ (which are related to the angle $\phi$), he presents the following differential forms:

$$dx = BT\text{ȣ}\, ds\ \cos\phi + T\, db\ \sin\phi$$

$$dy = BT\text{ȣ}\, ds\ \sin\phi - T\, db\ \cos\phi$$

To make these equations integrable, Euler introduces the simplification $\frac{1}{T\text{ȣ}} = \Theta$. He then proceeds to reduce the complex formulas into simpler forms, and he notes that for the integrability to be achieved, two specific quantities must depend solely on the parameter $b$.

The process here reflects Euler's approach to tackling the mathematical challenges in fluid mechanics by introducing new variables and simplifying complex expressions. The goal is to find a way to express the behavior of the fluid in a form that allows for integration, which is essential for obtaining meaningful physical solutions. However, Euler's method often involves these abstract transformations that, while mathematically sound, can be difficult to interpret physically.

*XLIII. Ex his iam poterimus diiudicare, num eiusmodi fluidi status, cuius resistentia perfecte sequatur regulam vulgarem, sit possibilis, et sub quibus conditionibus? id quod investigauisse operae erit pretium. Regula autem vulgaris postulat; ut sit* $u = \frac{\text{ȣȣ}}{C}$*; cum igitur hic posuerimus* $u = \text{ȣ}\cos\phi$*, fiet* $\text{ȣ} = C\cos\phi$*, existente C functione ipsius b tantum: hinc erit*

XLIII. From this, we can now determine whether such a state of fluid, whose resistance perfectly follows the common rule, is possible, and under what conditions? Investigating this will be worth the effort. However, the common rule requires that $u = \frac{\text{ȣȣ}}{C}$; and since we have assumed here that $u = \text{ȣ}\cos\phi$, it follows that $\text{ȣ} = C\cos\phi$, with C being a function of b alone. Hence, it will be ...

This passage explores the conditions under which a fluid's behavior, specifically its resistance, adheres to a common rule. The equation $u = \text{ȣ}\cos\phi$ is used to express velocity in terms of a parameter $\phi$, leading to a relationship where ȣ depends on $\cos\phi$ and a function $C(b)$, which is related to the parameter $b$. The author aims to understand the implications of this relationship for the fluid's motion.

**Author's note:** The following four paragraphs (XLIV, XLV, XLVI, XLVII) focus on deriving the shape of a body that conforms to the "common rule." In these sections, the velocity component perpendicular to the body's surface is expressed as u = ȣcosϕ and ȣ = Ccosϕ, where C is a function dependent solely on b. However, these paragraphs consist mainly of elaborate mathematical derivations, with numerous variable substitutions, that do not significantly advance the theoretical understanding. The resulting expressions are complex and difficult to interpret. As translating these sections would primarily involve reproducing intricate mathematical formulas, we encourage readers to consult the original text for full details. Instead, we will summarize the key results, including the equations for the streamlines and pressure, as follows:

$$x^2 = \frac{4}{C^2}\left(\frac{1}{C^2} - y\right);\ p = V - \frac{\frac{2c^2}{C^2}}{x^2 - \frac{2c^2}{C^2}}, \text{where c is a constant.}$$

*XLVIII. Iam ergo audacter pronunciare possumus, regulam resistentiae vulgarem exacte locum habere non posse, nisi quando figura corporis AEB fuerit parabolica, et singuli rivuli a e b, a' e' b' quoque sint inflexi secundum parabolas, quae cum illa parabola AEB, tam axem EF, quam focum F, habeant communem, unde et vasis extremam oram $\alpha\varepsilon\beta$ secundum similem parabolam formatam esse oportet. Cum igitur in reliquis casibus ominibus regula vulgaris a veritate aberret, resistentia quoque aliam sequetur legem, neque isti regulae erit consentanea. Quando ergo specie huius regulae nonnulli seducti putauerunt, fieri posse, ut corpus in fluido nullam resistentiam passurum moveatur, propterea quod actio fluidi im partem posticam deftruat vim in partem anticam exertam, et in fluidis terrestribus haec destructio a tenacitate prohiberi censeatur; iam manifestum est, hanc conclusionem nullo modo admitti posse. Quia enim corpus parabolicum AEB non utrinque terminatur, hic casus neutiquam ad resistentiae doctrinam traduci potest.*

Fig. 2.

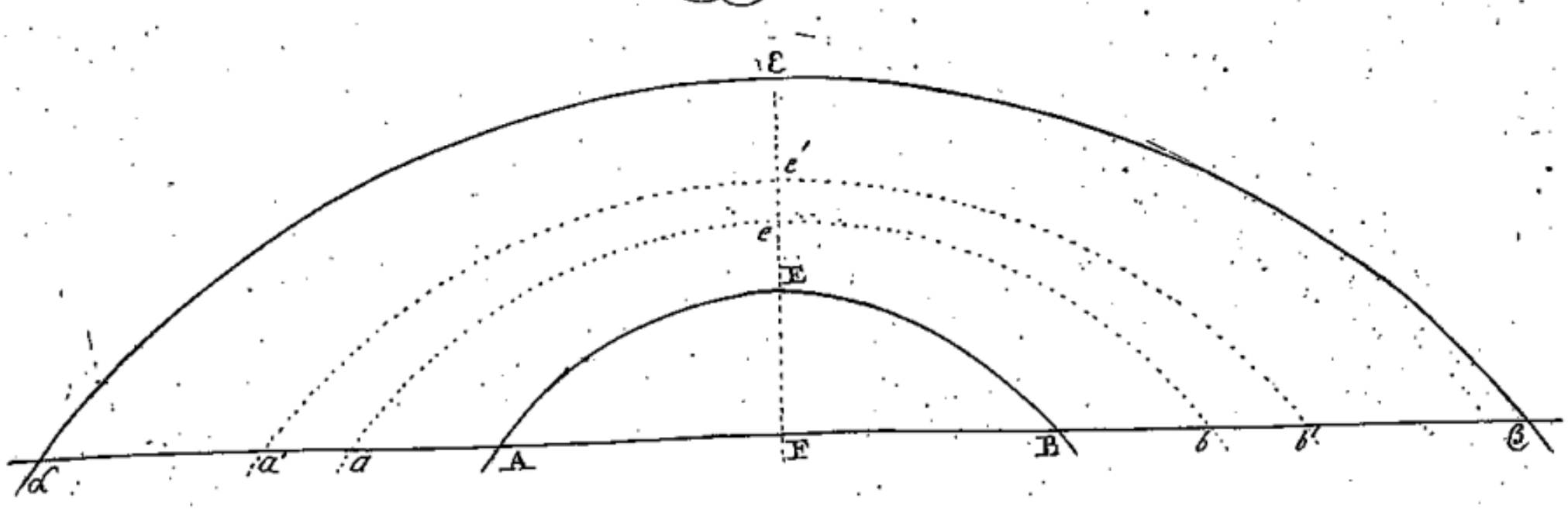

"XLVIII. Now, therefore, we can boldly declare that the common rule of resistance can only exactly apply when the shape of the body AEB is parabolic, and each stream a e b, a' e' b' is also curved according to parabolas that share both the axis EF and the focus F with the parabola AEB. Consequently, the outer edge of the vessel $\alpha\varepsilon\beta$ must also be shaped like a similar parabola. Therefore, in all other cases, the common rule deviates from the truth, and the resistance will follow a different law, not consistent with that rule. Hence, when some were misled by the appearance of this rule and believed that a body could move through a fluid without experiencing resistance—because the action of the fluid on the rear part cancels out the force exerted on the front—and that in earthly fluids this cancellation is prevented by viscosity, it is now clear that this conclusion cannot be accepted in any way. Since the parabolic body AEB is not terminated on both sides, this case cannot be applied to the doctrine of resistance."

This passage concludes that the common rule of fluid resistance is only accurate under very specific conditions—when the body and the fluid streams follow a parabolic shape. In all other cases, the rule does not apply, and resistance behaves differently. The author also refutes the idea that a body could move through a fluid without resistance based on the assumption that fluid forces cancel each other out. This refutation is significant in understanding the limitations of applying simple rules to complex fluid dynamics.